\documentclass[11pt]{article}
\usepackage{epsfig,amsfonts}
\usepackage[fleqn]{amsmath}
\usepackage{amsthm,amssymb}
\usepackage{graphicx}
\usepackage{hhline}
\usepackage{cite}
\usepackage{wick}
\def\be{\begin{equation}}
\def\ee{\end{equation}}
\def\bdm{\begin{displaymath}}
\def\edm{\end{displaymath}}
\def\bea{\begin{eqnarray}}
\def\eea{\end{eqnarray}}

\def\pp{\partial}

\newcommand{\rd}{\mbox{d}}
\newcommand{\ri}{\mbox{i}}
\newcommand{\re}{\mbox{e}}

\begin{document}

\begin{titlepage}
\begin{flushright}
RU-NHETC-2003-35\\
BONN-TH-2003-05\\
\end{flushright}

\vspace{1.5cm}

\begin{center}
\begin{LARGE}
{\bf Integrable Model of Boundary Interaction:}

\vspace{0.3cm}

{\bf The Paperclip}

\end{LARGE}

\vspace{1.3cm}

\begin{large}

{\bf S.L. Lukyanov}$^{1,2}$, {\bf E.S. Vitchev}$^1$ 

\vspace{.2cm}

{\bf and}
\vspace{.2cm}

{\bf A.B. Zamolodchikov}$^{1,2,3}$
\end{large}

\vspace{1.cm}

{${}^{1}$NHETC, Department of Physics and Astronomy\\
     Rutgers University\\
     Piscataway, NJ 08855-0849, USA\\

\vspace{.2cm}

${}^{2}$L.D. Landau Institute for Theoretical Physics\\
  Chernogolovka, 142432, Russia

\vspace{.2cm}

and

\vspace{.2cm}

${}^{3}$ Physikalisches Institut der Universit${\rm \ddot a}$t  Bonn\\
 Nu\ss allee 12,
D-53115 Bonn\\
Federal Republic of Germany}

\vspace{1.5cm}

\end{center}

\begin{flushleft}
\rule{4.1 in}{.007 in}\\
{December 2003}
\end{flushleft}
\vfill

\begin{center}
\centerline{\bf Abstract} \vspace{.8cm}
\parbox{11cm}{
We consider a model of $2D$ quantum 
field theory on a disk, whose bulk dynamics
is that of a two-component free massless Bose field ${\bf X} = (X,Y)$, and
interaction occurs at the boundary, where the boundary values $(X_B, Y_B)$
are constrained to special curve -- the ``paperclip brane''. The interaction
breaks conformal invariance, but we argue that it preserves integrability.
We propose exact expression for the disk partition function (and more general
overlap amplitudes $\langle\,{\bf P}\,| \,B\,\rangle$ of the boundary state
with all primary states) in terms of solutions of certain ordinary linear
differential equations.
}
\end{center}

\end{titlepage}
\newpage

\section{ Introduction} \label{secintro}

Lately, a class of $2D$ Quantum Field Theories (QFT) is attracting
attention from different points of view. These are the QFT which
are Conformal Field Theories (CFT) in the bulk, but have
non-conformal interactions at the boundary. In the
simplest setting, the bulk CFT is placed inside the disk $|
z|<R$ (where $( z, {\bar z})$ are standard complex
coordinates in the $2D$ space) and the non-conformal interaction
occurs at the boundary $|z| = R$. When the length scale
changes from short to long, such theories typically interpolate
between different conformal boundary conditions, and for that
reason they are often referred to as the boundary
Renormalization Group (RG) flows. Field theories of this type
are of interest in dissipative quantum mechanics, where the bulk CFT
theory plays the role \cite{Calan} of the Caldeira-Leggett quantum
thermostat \cite{Legett}. They are also extensively
used in studying the Kondo model and related 
problems\ \cite{afflecka,saleura}.
Certain interest to such theories
exists in the string theory\ \cite{witten,samson,moore}.
From the last point of view,
of special interest are the ``brane models'', where the bulk CFT
contains a collection of $N$ free massless scalar fields
${\bf X}( z,{\bar z})$ whose boundary values
${\bf X}_{B} \equiv {\bf X}|_{| z|=R}$ are constrained to
some hypersurface ${\bf \Sigma} \subset {\Bbb R}^{N}$ -- the
``brane''. Here and below we use the term ``brane'' loosely -- the
stringy on-shell brane must be a conformal boundary condition,
while we do not impose this requirement. The constraint
gives rise to the boundary interaction which in general breaks the
conformal symmetry: the shape of the ``brane'' changes with the
renormalization scale.
The RG flow equation can be computed perturbatively, in the limit
when the curvature of the brane is small\ \cite{Leigh, Rychkov}.

In this work we restrict attention to special model of this kind,
which contains two scalars ${\bf X} = ( X , Y)$, and the
brane is a one-dimensional curve ${\bf \Sigma} \subset {\Bbb R}^2$
in the two-dimensional target space. We fix normalizations of
these fields by writing down the bulk action
\bea\label{action}
{\cal A} = {1\over\pi}\,\int_{|z|<R} \rd^2z\ \big(\,
 \partial_{z}  X
\partial_{\bar z}  X+ \partial_{z}  Y
\partial_{\bar z} Y\, \big) \, ,
\eea
where $\rd^2 z = \rd{\rm x} \rd{\rm y}$. In this case, when
${\Sigma}$ is one-dimensional, the RG flow equation was
computed up to two loops in\ \cite{Rychkov}.
It turns out that this equation is
satisfied by the following scale-dependent curve
\bea\label{shape}
r\,\cosh\big( {\textstyle{X_B\over\sqrt{n}}}\big) -
\cos\big(  {\textstyle{Y_B\over\sqrt{n+2}}}\big)= 0 \,, \ \ \qquad
|Y_{B}| \leq \pi\, \sqrt{n+2}\ ,
\eea
where $n$ is real parameter independent of the RG energy scale $E$,
but the coefficient $r = r(E)$ ``flows'' 
with the scale $E$ according to
the
equation
\bea\label{flow}
\kappa = n\,  r^{\,n}\,\big(1 - r^2\big)\,,
\eea
where $\kappa$ is inversely proportional to $E$: $\kappa = 
\textstyle{E_{*}\over E}$. The
proportionality coefficient $E_{*}$ (the 
integration constant of the RG flow
equation) sets the ``physical scale'' for the model:
physical quantities, like the overlap amplitudes\ \eqref{exponent}
below,
will essentially depend on the dimensionless combination $E_{*} R$.
In what follows we always take the normalization scale $E$ equal to
$R^{-1}$, so that $\kappa$ in 
the left-hand side of\ \eqref{flow}\ coincides
with this combination,
\bea\label{kappadef}
\kappa = E_{*} R\,.
\eea
Eq.\eqref{flow} then relates $r$ to the radius $R$.
The curve\ \eqref{shape} is sketched in  Fig.1.

\begin{figure}[h]
\centering
\includegraphics[width=12cm]{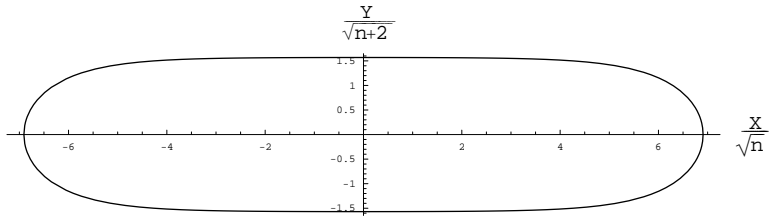}
\caption{The paperclip shape for $r=2\times 10^{-3}$. }
\label{fig-ampl}
\end{figure}

\noindent
It has a paperclip
shape, and we will refer to this theory as the ``paperclip-brane
model'', or simply the ``paperclip model''. At short scales, 
i.e. at $R \to 0$, the
coefficient $r$ becomes small, and the paperclip
grows long in the horizontal direction, with its length 
$\sim {2\over\sqrt{n}}\,\log\big({\textstyle{
1\over \kappa}}\big)$, while its width approaches
the constant $\pi\, \sqrt{n+2}$.
                                                                                
Strictly speaking, the two-loop paperclip solution\
\eqref{shape},$\,$\eqref{flow} applies
only when $n\gg 1$, for otherwise the curvature at the round ends of
the paperclip in Fig.1 is not small, and higher loops and
perhaps nonperturbative corrections should not be neglected. For
similar reason, Eq.\eqref{shape} can be taken literally only at
sufficiently small length scales, where $r$ is well below 1. For
larger length scales a 
nonperturbative description of the paperclip-brane
model is needed. At the same time, there are reasons to believe
that existence of consistent local boundary conditions is not
limited to the perturbative domain $n\gg 1$, and in this sense the
``paperclip'' model can be given appropriate nonperturbative
definition for all $n>0$. It is the problem we address in this work.

Nonperturbative description of a local boundary condition can be given
in terms of the associated boundary state\ \cite{nappi}.
The boundary state
$|\, B\, \rangle$ is a special vector
in the space of states ${\cal H}$ of
radial quantization of the bulk CFT which incorporates all the effect
of the boundary at $|z|=R$. In our case ${\cal H}$ is the usual space of
states of two-component uncompactified scalar,
\bea\label{fullspace}
{\cal H} = \int_{\bf P}\ {\cal F}_{\bf P} \otimes {\bar{\cal F}}_{\bf
P}\,,
\eea
where ${\cal F}_{\bf P}$ is the Fock space of two-component
right-moving boson with the zero-mode momentum ${\bf P}=
(P,Q)$.
The overlap $\langle\,  v\, | \, B\,  \rangle$
with
any $|\, v\, \rangle \in {\cal H}$ coincides, up to the factor
$R^{c/6-2\Delta_v}$, with the
(unnormalized)
expectation value
$\langle\,  {\cal O}_v (0,0)\, \rangle_{\rm disk}$ of
the field ${\cal O}_v (z,{\bar z})$ associated with
$|\, v\, \rangle$ through
the standard operator-state correspondence of the
bulk CFT. In particular, the overlap $\langle\, {\bf  P}\, |\,  B\,
\rangle$
with the Fock vacuum $|\,  {\bf P}\, \rangle$ relates to the
unnormalized one-point function of associated primary field
inserted at the center of the disk,
\bea\label{exponent}
\big\langle\,  \re^{\ri {\bf P}\cdot {\bf X}} (0,0)
\, \big\rangle_{\rm disk} =  R^{1/3-{{\bf P}^2/ 2} }\
\langle\, {\bf  P}\, |\, B\,  \rangle\,,
\eea
and $R^{1/3}\ \langle\, {\bf  0}\, |\,  B\,  \rangle$ is the disk
partition
function\footnote{We assume that
the exponential field in \eqref{exponent} is
defined according to
the usual CFT conventions \cite{Book}. The same will
apply below to all composite fields in the bulk of the disk. Namely,
unless
specified otherwise, all bulk composite fields appearing below are
understood as the normal-ordered products, taken with respect to the
 standard Wick paring $\overwick{ {X}{^\mu (z,\bar z)}
{X}{^\nu (w,\bar w)} }
{\contract{}{X} {X}{^\nu (w,\bar w}}
=  -\delta^{\mu\nu}\  \log|z-w|$.}.
When the boundary condition is not a conformal one, the overlaps depend
on the combination \eqref{kappadef}\,; 
to stress this fact we will use the
notation
\bea\label{partit}
\langle\, {\bf  P}\, |\,  B\,  \rangle =
Z(\, {\bf P}\, |\, \kappa\, )\,.
\eea

In this paper we propose exact expression for the function $Z(\, {\bf
P}\, |\, \kappa\, )$
in terms of solutions of certain ordinary linear differential equation.
Our
proposal is summarized in  Eqs.\eqref{diff}-\eqref{exactz}
in Section\ \ref{secseven1}\ below. The motivations and
evidence for this proposal 
are presented in turn in the main part of the
paper. Here we would like to outline the logical structure of our
arguments,
and their relation to known exact results in other models of boundary
interaction.
                                                                                
First, we are going to argue 
that the paperclip-brane model is integrable.
More precisely, we will argue that this model fulfills the paradigm of
boundary integrability as described in Ref.\cite{gz}.
We will show that the holomorphic sector of the bulk CFT
\eqref{action} contains special set of commuting local
integrals of motion $\big\{\, {\Bbb I}_s\,;\, s=1,3,5\ldots \big\}$, 
and that the
paperclip boundary constraint preserves the combinations
${\Bbb I}_s - {\bar{\Bbb I}}_s$, where ${\bar{\Bbb I}}_s$ are the
antiholomorphic
counterparts of ${\Bbb I}_s$. In the radial
quantization picture, 
where ${\Bbb I}_s$ are hermitian operators acting in
the
space ${\cal F}_{\bf P}$, such that
\bea\label{commute1}
[\, {\Bbb I}_s\, ,\, {\Bbb I}_{s'}\, ] =0
\eea
for all $s,s'$, the precise form of the last statement is that the
paperclip-brane boundary state $|\, B\, \rangle$ satisfies the equations
\bea\label{integrable}
(\, {\Bbb I}_s - {\bar{\Bbb I}}_s\, )\ |\, B\,  \rangle = 0
\eea
for all $s=1,3,5\ldots$\ . The first integral ${\Bbb I}_1$ essentially
coincides  with
$\textstyle{{\Bbb L}_0
\over R}$, where
${\Bbb L}_0$ is
the zero-mode Virasoro generator, and so the first of the
equations is simply the statement of the translational invariance of the
boundary condition \eqref{shape}. Few further operators of the
``paperclip
series'' ${\Bbb I}_3,\,  {\Bbb I}_5\ldots$ are described 
in Section\ \ref{secfive}\ below.
General explanations about the local meaning
of the equations \eqref{integrable} can be found in  Ref.\cite{gz}.

The equations \eqref{integrable} imply that the boundary state can be
written in terms of the simultaneous eigenstates $|\, \alpha\, ,\,  {\bf
P}\,  \rangle
\in {\cal  F}_{\bf P}$ of the operators ${\Bbb I}_s$ as follows,
\bea\label{bstatexp}
|\, B\, \rangle = \int_{\bf P}\,\sum_{\alpha}\,
B_{\alpha}({\bf P})\ |\, \alpha\, ,\,  {\bf P}\,  \rangle \otimes
\overline{|\,\alpha\, ,\,  {\bf P}\,  \rangle}\,,
\eea
(where the overlined states are the corresponding eigenstates of
$\{\, {\bar{\Bbb I}}_s\, \}$ in ${\bar{\cal F}}_{\bf P}$) with 
some c-number
coefficients $B_{\alpha}({\bf P})$.
Also, one can observe 
(see Section\ \ref{secfive}\ below) that the expressions for the
operators ${\Bbb I}_s$ do not involve any particular scale, hence the
eigenvectors $|\, \alpha\, ,\,  {\bf P}\, \rangle$ do not depend on the scale
$E_{*}$, while all the dependence on the parameter \eqref{kappadef} comes
through the coefficients $B_{\alpha}({\bf P})$ in \eqref{bstatexp}: 
$B_{\alpha}({\bf
  P}) = B_{\alpha}(\, {\bf P}\, |\, \kappa\, )$. 
This structure is typical for all 
integrable models of boundary interaction;
it suggests interesting general relation\ \cite{blz}\ between the integrable
boundary states and commuting 
operator families {\it ${\grave a}$ la} Baxter\ \cite{BXT}.
Assuming that the integrals of motion ${\Bbb I}_s$ are
``resolving'',
i.e. that for generic ${\bf P}$ all eigenspaces of $\{\, {\Bbb I}_s\, \}$ in
${\cal
  F}_{\bf P}$ are one-dimensional (which seems to be the case), one can
associate to the state \eqref{bstatexp} an operator ${\Bbb B}(\kappa)$
( the ``boundary-state operator'') acting in the space
$\int_{\bf P}\,{\cal F}_{\bf P}$,
\bea\label{boperator}
{\Bbb B}(\kappa) =
\int_{\bf P}\,\sum_{\alpha}\, B_{\alpha}(\, {\bf
P}\, |\, \kappa\, )\ \, |\, \alpha\, ,\,  {\bf P}\, \rangle\, 
\langle\,  {\bf P}\, ,\, \alpha\, |\, .
\eea
In writing \eqref{boperator} we have also assumed that the
eigenstates $|\, \alpha\, ,\,  {\bf P}\, \rangle$ are chosen orthonormal,
\bea\label{orthonormal}
\langle\, {\bf P}\, , \alpha\,  |\, \alpha'\, ,\,  {\bf P}'\, \rangle =
\delta_{\alpha,\alpha'}\ \delta^{(2)}({\bf P} - {\bf P}')\,.
\eea
Obviously, the operators ${\Bbb B}(\kappa)$ with different values of the
``spectral parameter'' $\kappa$ commute,
\bea\label{bcomm}
[\, {\Bbb B}(\kappa)\, ,\,  {\Bbb B}(\kappa')\, ] =0\,,
\eea
in suggestive similarity with Baxter's operators. And indeed, it was
observed in\ \cite{blz} that the
appropriately defined Baxter's operators of
minimal
CFT can be identified with the boundary-state operators associated
with
certain integrable boundary conditions, and this relation seems to be of a
general nature\ \cite{BLZb,baza,Tsvelik}.
Two more observations about Baxter's operators of
CFT will be important for our arguments.
                                                                                
First, not only the operator ${\Bbb B}(\kappa)$ commutes with the local
integrals $\{\, {\Bbb I}_s\, \}$, but in some cases it admits useful
asymptotic $\kappa\to\infty$ expansion in terms of these integrals.
Assuming
that the index $s$ labeling the local integrals ${\Bbb I}_s$ coincides
with
their scale dimensions, this expansion can be written as
\bea\label{bexpan}
\log{\Bbb B}(\kappa)\ \simeq\  \log{\Bbb B}_{IR}-  
\sum_{s}\,{{C_s}}\ {{\Bbb
  I}_s\over E_{*}^{s}}\ ,
\eea
where 
${\Bbb B}_{IR}$ is the boundary state operator associated with
the infrared fixed point, and
$E_{*}$ relates to $\kappa$ as in \eqref{kappadef}. The
dimensionless constants $C_s$ depend only 
on accepted normalizations of the
integrals of motion ${\Bbb I}_s$. 
In all known examples, the asymptotic  expansion 
\eqref{bexpan} is valid when the boundary flow associated with
${\Bbb B}(\kappa)$ ends in the ``trivial'' infrared fixed point, the one
which
in  Cardy's classification\ \cite{Cardy}\ corresponds 
to the identity primary
state; otherwise, the $\kappa\to\infty$ expansion of $\log {\Bbb
B}(\kappa)$
involves also non-local integrals of motion of 
fractional scale dimensions\ \cite{BLZb}.
Conversely, when an integrable boundary interaction flows to the
``trivial''
fixed point, the structure \eqref{bexpan} is natural to expect, because
the
``trivial'' conformal boundary condition features no local boundary fields
but
the descendents of the identity operator\ \cite{Cardy}.
                                                                                
Another important observation belongs to Dorey and Tateo, who have shown
that, in the case of minimal CFT, the vacuum-vacuum matrix elements of
Baxter's operators (i.e. the overlap amplitudes analogous to
\eqref{exponent})
can be written down explicitly, in terms of solutions of certain ordinary
linear differential equation\ \cite{toteo}\ (see also\ \cite{blzz}).
Albeit still rather
mysterious, this relation has proven to be rather general and very
useful.
The differential equations have been identified for quite a few different
classes of  Baxter's operators, 
where the relation can be either proven\ \cite{blzz,baza}\ or
verified against known ``TBA systems'' of functional relations. In our
case, for the paperclip model, 
the associated TBA system (which is related to the TBA system of the
``sausage'' sigma model\ \cite{sausage})\ is too complex to take it as
the starting point. On the other hand, a wealth
of data about the paperclip boundary state can be obtained through
perturbative analysis in the weak coupling domain, and by looking into
various
limiting cases of the model; 
Sections\ \ref{secthree}\ through\ \ref{secsix}\  below are devoted to
these tasks. Then one can simply
try to figure out the differential equation which would reproduce all these
data. This is the approach we have taken in this work. 
In  Section\ \ref{secseven}\ we
present special second-order differential equation,  
Eq.\eqref{diff}, and
propose the relation\ \eqref{exactz}\ expressing the amplitude
\eqref{partit} in
terms of certain solutions of that differential equation. 
Then we show that
this expression reproduces expected properties of the paperclip model in
all fine details.
                                                                                
The last thing we would like to mention before entering the details is the
circular limit of the paperclip. In the limit when $n\to\infty$ and $r$ is
taken
sufficiently close to 1, so that $
g^{-1} = n\,(1-r^2)$ remains finite, the
paperclip curve \eqref{shape} becomes a circle,
\bea\label{circle}
{\bf X}_{B}^2 ={\textstyle {1\over g}}\,,\ \ \ \ {\rm with}\ \ \ \ 
\kappa=g^{-1}\ \re^{-{1\over 2g}}\ ,
\eea
and we refer to this limiting case as the ``circular brane''
model. This model is important in view of its applications in
Condensed Matter Physics \cite{Kosterlitz, Eckern, Zaikin, Larkin},
in particular in relation to the 
problem of
Coulomb charging in quantum dots. The results for this 
special case were previously reported in 
Ref.\cite{SLAZ}.

\section{Elementary Paperclip}\label{sectwo}
                                                                                                                     
In this section we list few elementary properties of the paperclip
model, mostly to prepare the notations for the future discussion,
and to set the stage altogether.

\subsection{ Functional integral}\label{sectwo1}

The one-point function\ \eqref{exponent}
can be represented in terms of the functional integral as
follows,
\bea\label{path}
\big\langle\,  \re^{\ri {\bf P}\cdot {\bf X}} (0,0)
\, \big\rangle_{\rm disk}=\int\,
{\cal D}X\,{\cal D}Y\ \re^{\ri PX +
\ri QY}(0,0)\ \re^{-{\cal A}[X,Y]}\ ,
\eea
where ${\bf P}=(P,Q)$, and
the integration variables $X(z,{\bar z}),Y(z,{\bar z})$
are assumed to obey the constraint\ \eqref{shape} at the boundary
$|z|=R$.
In view of the compact nature of the paperclip\ \eqref{shape},
the integrand in \eqref{path} is bounded for any complex $P$ and
$Q$, and therefore the 
overlap\ $\langle\,  {\bf P}\, |\,  B\,  \rangle$\ in Eq.\eqref{exponent}
is expected to be an entire
function of these variables. For this reason it is useful to think
of the momenta $P,Q$ as the complex variables. For instance, when
$P$ and $Q$ are pure imaginary, some insight can be gained by
making a shift of integration variables
\bea\label{shift}
{\bf X}\to {\bf X}+\ri\, {\bf P}\ \log {\textstyle{{|z|}\over {R}}}\, ,
\eea
in the functional integral\ \eqref{path},
which brings it to the form
\bea\label{cyl}
\big\langle\,  \re^{\ri {\bf P}\cdot {\bf X}} (0,0)
\, \big\rangle_{\rm disk}=
R^{-{\bf P}^2/2 }\
\int\,{\cal D}{X}\,{\cal D}{Y}\ \re^{-{\cal A}[{X},{Y}]-{\cal
A}_{\rm B}[{X}_{B},{Y}_{B}]}\ ,
\eea
where the parameters $(P,Q)$
play the role of external fields
coupled to the boundary values ${\bf X}_B=
(X_B,Y_B)$
via the boundary action
\bea\label{baction}
{\cal A}_B = -  \oint_{|z|=R}{ \rd z\over 2\pi
z}\ \, \big( PX_{B}+QY_{B} \big) (z)\, .
\eea

\subsection{Topological sectors and instantons}\label{sectwo2}

The functional integral \eqref{path} is taken over all field
configurations
${\bf X}(z,{\bar z})$ inside the disk satisfying the paperclip
constraint\ \eqref{shape}\ at the boundary. Since topologically the
paperclip \eqref{shape}\ is a circle, the configuration space splits into
the topological sectors characterized by the integer-valued
winding number $w \in {\Bbb Z}$. This is the number of times the
boundary value ${\bf X}_{B}$ winds around the
paperclip when one goes around the circle $|z|=R$. As usual,
existence of the topological sectors allows one to generalize the
paperclip model by introducing the topological $\theta$-angle, so
that the contributions from different sectors are weighted with
the phase factor $\re^{\ri\theta w}$. The boundary state $|\, B\,
\rangle$ would then acquire the $\theta$-dependence. In this work
we do not attempt to study the full $\theta$-dependent paperclip
model, restricting attention to the case $\theta=0$. Nonetheless
it is important that even in this case the partition function
$Z(\, {\bf P}\, |\, \kappa\, )$ can be written as a sum
\bea\label{theta}
Z(\,{\bf P}\,  |\, \kappa\,) = \sum_{w\in {\Bbb Z}}\,Z^{(w)} (\, {\bf P}\,
|\, \kappa\,)
\, ,
\eea
of the contributions from all the winding sectors.

The decomposition\ \eqref{theta} is most
useful in the semiclassical domain,
where the contributions from different topological sectors are easier
to sort out. The semiclassical approximation is valid when the overall
size of the paperclip\ \eqref{shape}\  is sufficiently large. This is the
case
when $n\gg 1$  and $\log\big(
{\textstyle{1\over r}}\big)$ is not too small, so that $r^n\ll
1$. Note that according to\ \eqref{flow}\ one has to have sufficiently
small
$R$ in order to meet the last condition; in other words the
semiclassical domain corresponds to large $n$ and sufficiently small
length scales. In the leading semiclassical approximation the
contribution from given topological sector is dominated by the
classical solutions minimizing the action\ \eqref{action} in
that sector -- the
instantons. It is not difficult to find the instanton solutions.
At large $n$ one should not distinguish between $n$ and
$n+2$ in\ \eqref{shape}, and the action\ \eqref{action} can be written as
\bea\label{classaction}
{\cal A}_{\rm class} = {n\over{2\pi}}\ \int_{|z|<R}\rd^2z\
{\partial_zU\partial_{\bar z}U^*+
\partial_{\bar z}U\partial_{z}U^*\over U U^*}\ ,
\eea
where the complex-valued fields,
\bea\label{Ufield}
U=\re^{(X+{\ri}Y)/ \sqrt{n}}\, ,\ \ \ \ \ \
U^*=\re^{(X-{\ri}Y)/ \sqrt{n}}\ ,
\eea
satisfy the paperclip constraint which in this context is best written
in the form
\bea\label{neshape}
\big( r U_B - 1 \big)\big( r U^*_B - 1 \big) = 1-r^2\ ,
\eea
with $U_B\equiv U|_{|z|=R}$.
In a given topological sector $w$ the action\ \eqref{classaction}
is bounded from
below, ${\cal A}_{\rm classical} \geq |w|\ A/2\pi$, where
$A=2\pi\,  \log\big({\textstyle{1\over  r^n}}\big)$ 
is the area of the paperclip\  \eqref{shape}.
This
bound is established by standard manipulations\ \cite{Belavin},
with the winding
number $w$ written in the form
\bea\label{winding}
w = {n\over{A}}\,\int_{|z|<R} \rd^2 z\ {\partial_{z}
 U {\partial}_{\bar z}U^* -
{\partial}_{\bar z}U \partial_{z} U^*\over U U^*}\ .
\eea
For $w\geq 0$ one finds
\bea\label{bound}
{\cal A}_{\rm class}^{(w)} - {{w A}\over{2\pi}} = {n\over\pi}\,\int
\rd^2 z\ \Big|{{\partial_{\bar z}}U\over U}\Big|^2 \geq 0\,;
\eea
this equation also shows that the minimum is achieved when $U = U(z)$
is a holomorphic function of $z$ inside the disk, satisfying the
constraint\ \eqref{neshape} at the boundary. Since
Eq.\eqref{neshape}\  defines a circle, such functions exist and
have the form 
\bea\label{instantons}
U(z) = {1\over r} - {\sqrt{1-r^2}\over r}
\ \ \re^{\ri\phi}\, \prod_{k=1}^{w}\,{{(z -
a_k)\, R}\over{R^2 -a_{k}^* \,z}}\ ,
\eea
where
$a_k$ are arbitrary complex numbers
inside the disk, $|a_k| < R$, and $\phi$ is an arbitrary 
phase\footnote{Eq.\eqref{instantons} generalizes the instanton solutions 
of the ``circular-brane'' model, which were found in \cite{Korshunov,
Nazarov}.}\,.
One easily checks that $U(z)$ defined by this equation is indeed
holomorphic for $|z|<R$ as long as $r <  1$. Similar expression,
with $z$ replaced by $\bar z$, can be obtained in the case of negative
$w$. The parameters $a_i$ and $\phi$ are the moduli of the
$w$-instanton solution. Roughly speaking, the deviations $R-|a_i|$
represent sizes of the individual instantons, while the phases of
$a_i$ correspond to their locations along the boundary circle $|z|=R$.
For any $w$-instanton configuration
\bea\label{shysy}
\exp\big(-{\cal A}_{\rm class}^{(w)}\, \big) = r^{n|w|}\ ,
\eea
i.e. the contributions from the sectors with $w\neq 0$ are suppressed
by  small factors $r^{n|w|}$.
                                                                                                                     
Nonetheless, instantons in the paperclip model give rise to a subtle
effect generally known as the ``small-instanton
divergence'' (see\ \cite{Frolov}). 
Evaluation of the instanton contributions involves
integration over the moduli space, and this integral diverges
logarithmically when the parameters $a_i$ approach the circle
$|a_i|=R$. The divergence is clearly seen in the one-instanton
calculation, see  Section\ \ref{secthree2}\ below. This divergence can not be
absorbed into the renormalization of the parameters of the curve
\eqref{shape};
one has to introduce an unusual boundary counterterm proportional to
$r^n$.
Similar phenomenon is known to exist in the  $O(3)$ sigma 
model\ \cite{Frolov,Lusher}
and in
circular
case of our model (see e.g.\ \cite{SLAZ}\ and references therein). 
One important effect of the small-instanton divergence is that it
gives rise
to a weakly singular factor in the partition function \eqref{partit},
\bea\label{sfactor}
Z(\, {\bf P}\, |\, \kappa\, ) = \big(C\kappa\big)^{2\kappa}\ 
{\tilde Z}(\, {\bf
  P}\, |\, \kappa\, )\,,
\eea
where ${\tilde Z}$ admits normal perturbative (in the powers of $r$) and
``instanton'' expansion (in the powers of $\kappa$). The factor $C$ in
\eqref{sfactor} represents ambiguity which derives from the small
instantons --
it depends on specific regularization of the 
small-instanton divergence.
We will say more about this effect in 
Section\ \ref{secthree2}\ below. Here we just observe that the ambiguity
affects the extensive energy of the paperclip boundary, i.e. the
coefficient $f$ in the asymptotic of $Z$ at large $R$,
\bea\label{extensive}
\log Z(\, {\bf P}\, |\, \kappa\, ) 
\to  f (\,{\bf P}\,)\,\kappa \qquad {\rm as} \quad
\kappa \to\infty\,.
\eea
In what follows we assume that the ambiguity is fixed by imposing the
normalization condition
\bea\label{fixamb}
f(\,{\bf 0}\,) = 0\, ,
\eea
which is always possible to achieve by appropriate adjustment
of the constant $C$ in \eqref{sfactor}\,.

\section{ Semiclassical Paperclip}\label{secthree}

Here we evaluate leading semiclassical
contributions to the overlap\ \eqref{exponent}\ by direct
calculation of the functional integral\ \eqref{path}\ in the
saddle-point approximation. This will give some intuition about its
structure. We will consider contributions from the
zero- and the one-instanton sectors only.

\subsection{ Zero winding}\label{secthree1}

Let us first assume that the parameters $P,Q$ are sufficiently small
so that vertex insertion in \eqref{path}\ is ``light'', i.e. it 
has no appreciable effect on the
saddle-point configurations. We write
\bea\label{kl}
(P,Q) ={\textstyle \frac{2}{\sqrt{n}}}
 \ (p,q)
\eea
and assume that $p,q\sim 1$.
Then, at zero winding the action is minimized by the trivial classical
solutions ${\bf X}(z,{\bar z}) = {\bf X}_0$, where ${\bf X}_0 =
(X_0 ,
Y_0)$ is an arbitrary point on the paperclip\ \eqref{shape}. Therefore
\bea\label{mini}
Z^{(w=0)}_{\rm class} = \int_{\rm
paperclip}\rd{\cal M}({\bf X}_0)\ \re^{2\ri (p{X_0} +
{q Y_0})/\sqrt{n}}\ ,
\eea
where the integration measure $\rd{\cal M}({\bf X}_0)$ is determined by
integrating out  fluctuations around the classical solution in
the Gaussian approximation. Of course there is no need of actually
evaluating this Gaussian functional integral to figure out the
answer. If one writes ${\bf X}(z,{\bar z}) = {\bf X}_0 + \delta
{\bf X}(z,{\bar z})$, and splits the fluctuational part $\delta
{\bf X}$ into the components normal and tangent to the paperclip at
the point ${\bf X}_0$, the components are to satisfy the Dirichlet and
the Neumann boundary conditions, respectively. Therefore
one just has to take the product of the known (see e.g. Appendix to
\cite{saleur}) disk partition functions with these two boundary
conditions. As the result, $\rd {\cal M}({\bf X}_0)=
{\textstyle{g_D^2\over 2\pi}}\, \rd\ell({\bf X}_0)$, 
where $\rd\ell({\bf X}_0)
=\sqrt{(\rd X_0)^2 + (\rd Y_0)^2}$ is the
length measure of the paperclip, and
$g_D = 2^{-{1\over 4}}$ is  the ``boundary degeneracy'' 
\cite{affleck} associated with
the Dirichlet conformal boundary condition\footnote{The definition is as 
follows: $g_D = \langle\,  P\, |\, B_D\, \rangle$, 
where $|\,B_D\,\rangle$ is
the boundary state of {\it uncompactified} boson $X$ with the 
Dirichlet boundary
condition $X_B=0$, and the primary states $|\, P\,  \rangle$ are
delta-normalized, $\langle\,P\,\mid \,P'\,\rangle = \delta(P -P')$.}.
Certainly, this
argument ignores the curvature of the brane at ${\bf X}_0$, but its
proper account leads to the one-loop part of the renormalization of
the parameter $r$, as described by Eq.\eqref{flow}
(see \cite{Leigh}).
                                                                                                                     
Of course there are many ways to evaluate the integral\ \eqref{mini},
but the
instructive one is to change to the variable
\bea\label{variab}
U_0=\re^{(X_0+{\ri}Y_0)/ \sqrt{n}}
\ \ \,, \qquad U^*_0 = {{U_0-r}\over{rU_0-1}}\,,
\eea
where the expression
for $U_{0}^*$ follows from \eqref{neshape}; 
this brings\ \eqref{mini}\ to the form
\bea\label{integra}
&Z^{(w=0)}_{\rm class}=&
g_D^2\ \sqrt{n\, (1-r^2) }\times\\ &&
\oint{\rd U_0\over  2\pi\ri}\
U_0^{-{ \frac{1}{2}}+q+\ri p}\ (rU_0 - 1)^{-{1\over 2}+
q-\ri {p}}\ (U_0 - r)^{-{1\over 2}-
q+\ri {p}}\ .\nonumber
\eea
The integration here is taken over the circle  $|rU_0-1|=\sqrt{1-r^2}$
(see Eq.\eqref{neshape})
which goes around
the branching points $U_0=r$ and $U_0={\textstyle{1\over r}}$. 
By closing the contour on
the branch cut between these two points the integral is brought to a
standard representation of the hypergeometric function,
\bea\label{semiclas}
Z^{(w=0)}_{\rm class} =
g_D^2\ \sqrt{n\,(1-r^2)}\ r^{
-2\ri p}\ _2 F_1 \big({\textstyle \frac{1}{2}}+q-
\ri\,  p, {\textstyle {1\over 2}}-q-\ri\, p
, 1; 1-r^2\big)\,.
\eea
It is also instructive to rewrite this result in another
form. The integration contour can be represented as the combination of
two contours, both going from $U_0=0$ back to $U_0=0$, one around the point
$U_0=r$, and another around $U_0={\textstyle{1\over r}}$. 
Again the hypergeometric integrals
emerge, and one obtains
\bea\label{newsemi}
Z^{(w=0)}_{\rm class} = B_{\rm class}(p,q)\  F_{\rm
class}(p,q\, |\, r) + B_{\rm class} (-p,q)\  F_{\rm
class}(- p,q\, |\, r)\, ,
\eea
where
\bea\label{semirefl}
B_{\rm class}(p,q) = 
{g_D^2\ {\sqrt{n}\ r^{2\ri p}\ \Gamma(-2\ri\, p)}\over
{\Gamma( {\textstyle \frac{1}{2}} - q - \ri\, p)
\Gamma( {\textstyle \frac{1}{2}} + q -
\ri\, p)}}\ ,
\eea
and
\bea\label{artush}
F_{\rm class}(p,q\,|\,r) =
\sqrt{1-r^2}\  _2 F_1 \big({\textstyle \frac{1}{2}}+q+
\ri\,  p, {\textstyle \frac{1}{2}}-q+\ri\,  p
, 1+2\ri\, p; r^2\big)\ .
\eea
In this form the nature of singular behavior at $r\to 0$ is more
explicit. Let us make a (trivial) observation that the poles of
$B_{\rm class}(p,q)$ in the variable $p$ in the first term in\
\eqref{newsemi}
are cancelled by the poles in the higher terms of the hypergeometric
series $ F_{\rm class}$ in the second term, and vice
versa, in agreement with the statement that the full partition
function is an entire function of $P$ and $Q$.
                                                                                                                     
Although the above result was derived under the assumption that $P,Q$
are small, it needs little fixing to become valid for much 
larger values
of these parameters. When $(P,Q)$ become as large as
$\sqrt{n}$ the vertex insertion in\ \eqref{path}\ is ``heavy'', i.e. it 
must be treated as a part
of the action, and it affects the saddle-point analysis. In this case
the form \eqref{cyl}\ of the functional integral is more convenient.
The saddle-point configuration(s) is still a constant field,
${\bf X}(z,{\bar
z}) = {\bf X}_0$ (this time we are talking about the shifted field,
Eq.\eqref{shift}), but now ${\bf X}_0$ is not an arbitrary point on
the curve\ \eqref{shape},
but has to extremize the boundary action\ \eqref{baction},
\bea\label{jsxs}
{\cal A}_B[\, {\bf X}_0\, ]= -\ri\, {\bf P}\cdot{\bf X}_0\ .
\eea
There are two solutions of the saddle-point equation. They are easily
visualized in the case of pure imaginary $P$ and $Q$, when the associated
saddle-point field ${\bf X}$ is real valued: the points ${\bf X}_0$
are located at two opposite sides of the paperclip\ \eqref{shape}, one
corresponding to the minimum, and another to the maximum of the action
\eqref{jsxs}.
Of course the minimum dominates, but it is useful to keep both
in mind.
The saddle-point action plus the Gaussian integral over the
constant mode produce nothing else but the $p,\, q\to\infty$ asymptotic
form of the expression\ \eqref{semiclas} -- after all, this asymptotic of
the
integral\ \eqref{integra} is
controlled by the same saddle points. One can observe
that if one splits the constant-mode integration into two parts, as
was suggested in deriving\ \eqref{newsemi},
the parts receive contributions
from different saddle points -- one from the ``minimum'' and one from
the ``maximum'' one; this is why one of
the terms in\ \eqref{newsemi}
becomes negligibly small at $\Im m\,P \sim \sqrt{n}$. What makes
the difference at $(P,Q) \sim \sqrt{n}$ is the proper treatment of
non-constant modes. One writes
\bea\label{ksxusiu}
{\bf X}(z,{\bar z}) = {\bf X}_0 + {\bf t}_0 \ \delta X_t (z,{\bar z}) +
{\bf n}_0\ \delta X_n (z,{\bar z})\,,
\eea
where ${\bf X}_0$ is the position of the saddle point on the paperclip, and
${\bf t}_0$ and ${\bf n}_0$ are unit vectors tangent and normal to the
paperclip at this point. Then for small $\delta X_t$ the boundary
constraint\ \eqref{shape}\  reads
\bea\label{deltaxn}
\delta X_n =- {{r}\over 2 \sqrt{n(1-r^2)}}\,
\cosh\big( \textstyle{{X_0}\over \sqrt{n}}\big)\  \delta
X_{t}^2 + O(\delta X_{t}^3)\ ,
\eea
and, up to  higher-order terms, the boundary action
\eqref{baction} can be written as
\bea\label{bactionb}
{\cal A}_B= A_{B}[\, {\bf X}_0\, ] \mp {P_r \over {2\sqrt{n}}}\,
\oint \,{{\rd z}\over{2\pi z}}\,\delta X_{t}^2 \ ,
\eea
with the coefficient $P_r$ proportional to the normal component of the
``external field'' $ {\bf P}$ times the curvature in\ \eqref{deltaxn}.
Explicitly,
\bea\label{nub}
P_r =  \sqrt{P^2 + r^2\,Q^2\over 1-r^2}\ ,
\eea
where we assume that the branch of the square root is chosen in such a
way that $\Im m\,P_r \geq 0$; then the sign plus (minus) in
\eqref{bactionb}
applies to the ``minimum'' (``maximum'') saddle point.
Thus, while to the leading approximation the normal component $\delta
X_n$ still can be treated with the Dirichlet boundary condition, the
``boundary mass'' term in\ \eqref{bactionb}\ has to be taken to account
in evaluating the contribution from $\delta X_t$.
Note that for ${\textstyle {P_r\over \sqrt{n}}} 
\sim 1$ (which we assume) the
energy scale associated with this ``boundary mass'' is $\sim R^{-1}$,
so that the use of the renormalized paperclip parameter $r$ 
defined as in\ \eqref{flow}
is still appropriate. Using the well known boundary amplitude of the free
field
with quadratic boundary interaction \eqref{bactionb}\ \cite{Witten},
one finds that
Eq.\eqref{newsemi} would apply
to the case of $(P,Q)\sim \sqrt{n}$ as well if
one
puts corresponding additional factors in the two terms in\
\eqref{newsemi},
i.e. replaces  $B_{\rm class}(p,q)$  there by
\bea\label{shsdy}
{\tilde B}_{\rm class}(P,Q) = {B}_{\rm class}\big(\textstyle{\sqrt{n}
\over 2}\, P\, , \textstyle{\sqrt{n}\over 2}\, Q\big)\
\Gamma\big(1 -
 \textstyle{\ri P_r\over \sqrt{n}}
\big)\ ;
\eea
of course in this 
case one can use the $p,\, q \to\infty$ asymptotic forms of
the factors \eqref{semirefl}
and \eqref{artush}.

\subsection{  One-instanton sector}\label{secthree2}

Here we consider only the case of the ``light insertion'', i.e. assume
that
$P$ and $Q$ are of the order of ${\textstyle{1\over \sqrt{n}}}$, as 
in\ \eqref{kl}. Then the
saddle point configurations are the instanton solutions
\eqref{instantons}.
All the instanton solutions with $w=1$ are generated by 
${\Bbb S}{\Bbb U}(1,1)$ 
maps from
the ``canonical'' solution $U_0 (z)$:
\bea\label{oneinst}
U(z) = U_0 \Big({\textstyle
\re^{\ri\phi}\, {{(z-a) R^2}\over{R^2 - a^{*}\,z}}}\Big)\,,
\qquad
U_0 (z) = {1\over r} - {\sqrt{1-r^2}\over r}\,{z\over R}\ ,
\eea
with the moduli $a$, $a^*$ and $\phi$ appearing as the parameters of the
map.
The saddle-point contribution from the sector $w=1$ to the
functional integral \eqref{path} is the integral
over the moduli space,
\bea\label{modulint}
Z^{(w=1)}_{\rm class} =
r^{n}\,\int\,\rd{\cal M}(a,a^*,\phi)\ \big(U_0\big)^{\ri p+q}\,
\big(U^{*}_0\big)^{\ri p-q}\,,
\eea
where the factor $r^{n}$ is due to the saddle-point 
action\ \eqref{shysy}
and
$U_0$ stands for the solution \eqref{oneinst} evaluated at the center of
the
disk, $U_0 = U(0)= U_0 (-\re^{\ri\phi}\,a)$. The integration measure
$\rd{\cal M}(a,a^*, \phi)$ should
be computed as usual by integrating out the Gaussian fluctuations around
the
instanton
solution. The calculations are analogous to those in\ \cite{Larkin}\ but
tedious, we
will
present them elsewhere. The resulting form of the expression
\eqref{modulint} is
\bea\label{modulinte}
r^{n}\, \int_{|a|<R}\,{\rd^2 a\over 2\pi}\
{ g_D^2\, \big(n\,(1-r^2)\big)^{3\over 2}
\over r\, (R^2 - aa^*)}\ \int_{0}^{2\pi}\
{{\rd\phi}\over{2\pi}}\, \big(U_0 \big)^{-{1\over
    2}+\ri p+q}\,\big(U_{0}^{*}\big)^{-{1\over 2}+\ri p-q}\ ,
\eea
where $\rd^2 a=
\rd(\Re e\, a)\, \rd(\Im m\, a)$, and
again $U_0$ stands for $U(0) = U_0 (-\re^{\ri\phi}\,a)$.

The small-instanton problem is explicit 
in \eqref{modulinte}. The integral
over
$|a|$ diverges when $|a|\to R$. This 
limit corresponds to small instantons --
when $|a|$ is close to $R$ the function $U(z)$ in \eqref{oneinst} is
almost
constant (equal to the ``background'' value $U_0 = U(0)$) everywhere in
the disk except for a
small neighborhood of the point $z=a$, where the instanton is localized.
The
integral\ \eqref{modulinte}\ requires regularization. 
Any regularization is
essentially an instruction to suppress contributions from the instantons
of
the sizes $\Delta \leq \Lambda^{-1}$, where $\Lambda$ is the ultraviolet
cutoff of the theory. The integral
\eqref{modulinte} thus produces a $\Lambda$-dependent term,
\bea\label{divterm}
Z_{\rm class}^{(w=1)} =
\kappa\,\log\big(\Lambda\,{R}\big)\,\,Z^{(w=0)}_{\rm class} + {\rm finite}\ .
\eea
Similar divergences appear in the multi-instanton sectors, and by the
usual
arguments (which also take into account the ``instanton -- anti-instanton''
configurations) the divergent contributions exponentiate, leading to the
overall $\Lambda$-dependent factor
\bea\label{divexp}
\exp\big(2\kappa\,\log (\Lambda\,R)\big)
\eea
in the partition function $Z(\, P,Q\, |\,\kappa\, )$. The term
$2\kappa\,\log\Lambda$ in this exponential can be absorbed by 
local boundary counterterm
\bea\label{counterterm}
2\ri\, E_{*} R\,
\oint_{|z|=R}\,{\rd z\over{2\pi z}}\ \log\big(
{\textstyle{\Lambda\over CE_*}}\big)\,,
\eea
leaving behind the singular factor shown in \eqref{sfactor}. Note that the
counterterm \eqref{counterterm} has singular dependence on the paperclip
parameter $r$ -- that is what makes it distinct from usual perturbative
counterterms.

All possible regularizations of the integral \eqref{modulinte} essentially
differ in the notion of the ``size'' $\Delta$ of a small instanton. Here
we assume a natural definition
\bea\label{smallsize}
\Delta = {{R^2 - aa^*}\over{R\,|U_0|}}\, ,
\eea
which respects all evident symmetries of the paperclip
\eqref{shape}\footnote{Of course this choice is not unique. Generally
speaking, any choice different by a factor, $\Delta \to C(X_0,
  Y_0)\,\Delta$, where $X_0,Y_0$ is a ``background point'' related to
$U_0$ as
in \eqref{variab}, is acceptable as long as $C(X_0,Y_0)$ respects the
symmetries. Obviously, this ambiguity is related to the possibility to
add  counterterm \eqref{counterterm} with $C$ being a local function of the
boundary fields, $C = C(X_B,Y_B)$. Therefore in our case (unlike the
circular-brane model where the symmetry restricts the ambiguity to a
constant $C$) regularization
of the small-instanton divergence does have certain effect on the
physical
content of the theory. From this point of view \eqref{smallsize} is
rather a
part of our definition of the paperclip model. We will see that this
choice
is consistent with the integrability of the model.}.
With this choice, the easiest way to handle the integral \eqref{modulinte}
is to insert regularizing factor
$\Delta^{2\epsilon}$
into the integrand. Then, after subtracting
${\textstyle {\kappa\over \epsilon}\
(C E_{*})^{-\epsilon}}\ Z_{\rm
  class}^{(w=0)}$ (which is the job of the counterterm
\eqref{counterterm}),
the limit $\epsilon\to 0$ can be taken. 
The integration can be handled in
a way similar to that in the previous subsection, 
and the result for the
finite
part in \eqref{divterm} is
\bea\label{zinst}
&&{\textstyle{1\over 2}}\ 
g_D^2\  \big(n\, (1-r^2)\big)^{3\over 2}\ r^{n-2\ri p}\times\\
&&\ \ \ \ \ \ \ 
{\rd\over{\rd\epsilon}}\ {}_2F_1\big({\textstyle{1\over 2}} + q - \ri p
+\epsilon, {\textstyle{1\over 2}} - q -\ri p + \epsilon, 1+2\epsilon;
1-r^2)\big)\big|_{\epsilon=0}\ . \nonumber
\eea
And similarly to \eqref{newsemi}, this part of the
one-instanton contribution can  
be written as the sum of two terms behaving at small $r$ as $r^{n\pm
  2\ri p}$ times a power series in $r^2$.

\section{ Ultraviolet Paperclip}\label{secfour}

As was mentioned in  Introduction, at short length scales the
parameter $r$ in \eqref{shape}
becomes small and the paperclip grows long
in the $X$-direction. 
In the limit $r\to 0$ it can be regarded as a
composition of two ``hairpins'', as depicted in 
Fig.2. 

\begin{figure}[h]
\centering
\includegraphics[width=10cm]{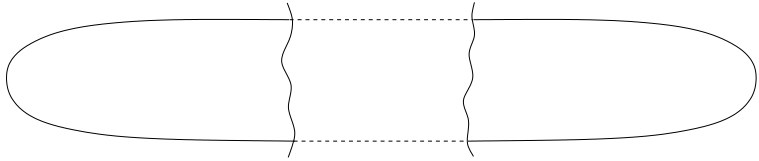}
\caption{The paperclip formed by a junction of two hairpins.}
\label{fig-ampla}
\end{figure}

Qualitatively, 
one expects that if $\Im m \,P$ is not too
small, the $R\to 0$ behavior 
of the overlap $\langle\, {\bf P}\, |\, B\,
 \rangle$
is controlled by one of 
the hairpins, at the left or at the right in
Fig.2, depending on 
the sign of $\Im m\,P$, with some crossover at
small $\Im m \,P$. It is useful therefore to describe the boundary
state of the hairpin brane in some details.

\subsection{ The hairpin}\label{secfour1}

For definiteness, we will discuss here the hairpin in the
left-hand side of  Fig.2; 
we call it the left hairpin. The right hairpin is 
the $X \to -X$ reflection of the left one. 
The (left) hairpin is described 
by the equation
\bea\label{hairshpe}
\textstyle{1\over 2}\, 
{\tilde r}\ \exp\big( \textstyle{-{X_B\over\sqrt{n}}}
\big) - \cos\big(\textstyle{Y_B\over\sqrt{n+2}}\big)=0
\, .
\eea
The hairpin is a brane in usual stringy sense -- the boundary
condition\ \eqref{hairshpe}\ is a conformal one. 
More precisely, the curve\ \eqref{hairshpe}
satisfies the RG flow equation with
\bea\label{RGtime}
{\tilde r} =\Big({\kappa\over n}\Big)^{1\over n}\, ,
\eea 
where $\kappa=E_*R$.
We have checked this up to two
loops, but it likely holds to all loops\footnote{As 
usual, the higher loop
  corrections are scheme dependent. 
The statement is that a scheme exists in
which Eq.\eqref{hairshpe} is valid to all orders in the loop
  expansion.}.  
The equation\ \eqref{RGtime}
shows that when the RG ``time'' $t = \log\kappa $ 
increases the hairpin 
``flows'' uniformly to 
the right with no change of shape. It can be
made into an RG fixed point in a precise sense of the word by an
appropriate redefinition of the RG transformation, 
namely by supplementing it with
a simple field redefinition $\textstyle{(X,Y) 
\to (X + {\delta t\over 
\sqrt{n}}, Y)}$. This corresponds to introducing a modified 
energy-momentum tensor
\bea\label{energymom}
T &=& - \partial_z  X \partial_z X - \partial_z Y \partial_z Y +
\textstyle{1\over\sqrt{n}}\,\partial_{z}^2 X\,,\\ \nonumber
{\bar T} &=& - \partial_{\bar z}
X \partial_{\bar z} X - \partial_{\bar z} Y \partial_{\bar z} Y +
\textstyle{{1\over\sqrt{n}}}\,\partial_{\bar z}^2 X\, ,
\eea
or, in a stringy speak, to introducing a linear dilaton $
D({\bf X}) =
\textstyle{{1\over \sqrt{n}}}\, X$ (for the problem at hand such
modification, in whatever speak, has no effect on the physical
content of the theory). With this, the boundary state 
$|\, B\, \rangle_{\subset}$ associated with the hairpin brane 
enjoys the conformal invariance in the usual form,
\bea\label{cftinv}
\big[\, 
z^2\ T(z) - {\bar z}^2 \ {\bar T}({\bar z})\, \big]_{|z|=R} \
|\, B\, \rangle_{\subset} = 0\,.
\eea
Moreover, the hairpin brane has an extended conformal symmetry
generated by higher spin currents -- the $W$-symmetry.
The generating currents $W_s (z)$ have spins $s=2, 3, 4\ldots$ 
(with $W_2
(z) = T(z)$), and can be characterized by the condition that they
commute with two ``screening operators'', i.e.,
\bea\label{comscreen}
\oint_{z}\rd w \ W_s (z) \,{\cal V}_{\pm}(w) = 0\,,
\eea
where
\bea\label{screen}
{\cal V}_{\pm} = \re^{\sqrt{n}\,X \pm \ri\sqrt{n+2}\,Y}\ ,
\eea
and the integration is over a small contour around the point $z$;
vanishing of the integral\ \eqref{comscreen}\ implies 
that the singular part of the
operator product expansion of $W_s (z)\,{\cal V}_{\pm}(w)$ is a total
derivative $\partial_w (\ldots )$. 
This condition fixes $W_s (z)$ uniquely
up to normalization and adding derivatives and composites of the
lower-spin $W$-currents. For instance, the first holomorphic current
beyond\ \eqref{energymom}\ can be written as
\bea\label{wwwcurrent}
\nonumber
W_3 &=& {3n+2\over 3}\ \big(\partial_z Y\big)^3 +
n\ \big(\partial_z X\big)^2 \partial_z Y +\\ 
&& {n\sqrt{n}\over 2}\ \partial^2_z
X\partial_z Y - {(n+2)\sqrt{n}\over 2}\ \partial X \partial^2_z Y +
{{n+2}\over 12}\ \partial^3_z Y\ ,
\eea
where the ambiguity in adding a term proportional $\partial_z T$ is
fixed by demanding that\ \eqref{wwwcurrent}\ is a conformal primary. 
One can notice that the current $W_3$ is antisymmetric under
reflection $Y\to -Y$, but has no symmetry with respect to $X \to -X$.
The higher currents $W_4,\,  W_5,\ldots $ can 
be found either by a direct
computation of the operator product expansions with the screening
exponentials\ \eqref{comscreen},
or recursively, by studying the singular parts of
the operator product expansions of the lower currents, starting with
$W_3 (z) W_3 (w)$ and continuing upward. Thus, the product $W_3 (z)
W_3 (w)$ contains singular 
term $\sim (z-w)^{-2}$ which involves, besides
the derivatives $\partial^2_z T$ and the composite
operator $T^2$, the new current $W_4$. Further operator products
with $W_4$ define higher $W$'s, etc. 
In this sense the $W$-algebra is
generated by the two basic currents $T$ and $W_3$. 
This $W$-algebra itself was
previously described in Ref.\cite{Bakas}. We also mention in
passing that when $n$ is a positive integer  the $W$-algebra thus
defined closes within $n-1$ lowest currents $W_2,\, W_3,\ldots W_n$
and coincides with the $WA_{n-1}$-algebra\ \cite{Zamol, FL}\ with 
the central
charge $c(n)=2+\textstyle{6\over n}$.

The hairpin boundary condition is invariant with respect to this
$W$-algebra, i.e.
\bea\label{wcftinv}
\big[\,
z^s\ W_s(z) - {\bar z}^s \ {\bar W}_s({\bar z})\, \big]_{|z|=R} \
|\, B\, \rangle_{\subset} = 0\, ,
\eea
for all the W-currents. To verify 
this statement by direct computation
with the constraint\ \eqref{hairshpe} is 
not that easy. We have checked that it
holds true in the classical limit $n\to \infty$, where the calculation
reduces to verification that the 
difference $z^3\, W_3 - {\bar z}^3\, {\bar
W}_3$ vanishes at the boundary in virtue of the classical equations of
motion of the hairpin model. In fact, in the classical case this
condition turns out to be very defining, in the sense that it fixes
the shape\ \eqref{hairshpe}\  of 
the hairpin uniquely, up to overall scale of the fields
$(X,Y)$ (in the classical case $n\to\infty$ one 
should not distinguish
between  $\sqrt{n}$ and $\sqrt{n+2}$ in\ \eqref{hairshpe}).  It is
plausible that  Eq.\eqref{wcftinv}\ can 
be verified order by order in the loop
expansion, but we did not perform any part of this task. Instead, we
have taken an attitude that  Eq.\eqref{wcftinv}, 
combined with  Cardy's
consistency condition\ \cite{Cardy}\ and 
qualitative properties of the 
hairpin\ \eqref{hairshpe},
can be taken as the definition of the quantum 
hairpin\footnote{In this approach 
one still has to check that this definition
agrees with the perturbative 
definition of the hairpin brane through 
Eq.\eqref{hairshpe}. We have checked that  Eqs.\eqref{amplit} and
\eqref{bounhair} below are in 
agreement with the one-loop calculations on the
disk with the boundary condition\ \eqref{hairshpe}.}. 

In view of\ \eqref{wcftinv}\ the hairpin 
boundary state can be written as the
combination of the Ishibashi 
states $|\, I_{\bf P}\, \rangle$ associated
with the above $W$-algebra\ \cite{Ishibashi}. 
An individual Ishibashi state
\bea\label{Ishib}
|\, I_{\bf P}\,  \rangle =
 \big[\, 1 + A_{\mu\nu}({\bf P})\ X^{\mu}_{-1}
\,{\bar X}^{\nu}_{-1} + \ldots\,  \big]\ |\,  {\bf P}\, \rangle
\eea
is the solution of the equations\ \eqref{wcftinv}\  in the Fock space
${\cal F}_{{\bf P}}\otimes{\bar{\cal F}}_{{\bf P}}$ with the zero-mode
momentum ${\bf P}=(P,Q)$. 
Here $X^{\mu}_{-k} = (X_{-k},\, Y_{-k})$ are the
bosonic creation operators 
associated with the oscillatory modes of
the fields $X^{\mu} = (X,Y)$. 
The solution is unique, 
i.e. the equations\ \eqref{wcftinv}\ determine
uniquely the coefficients $A_{\mu\nu}({\bf P})$, as well as all the
higher-level coefficients; for instance, the amplitudes $A_{\mu\nu}$ 
explicitly written
in\ \eqref{Ishib}
are (we rather write down certain linear combinations which
are shorter and more suggestive)
\bea\label{amplit}
&&A_{XX}-A_{YY} \pm  {{2\ri\,(n+1)}\over{\sqrt{n(n+2)}}}\,A_{XY} =
\\  \nonumber&&
\qquad  {{1+n}\over{1+n - 2\ri\sqrt{n}\,P}}\ \ {{1 + 2\ri\sqrt{n}\,P \mp
2\sqrt{n+2}\,Q}\over{ 1 - 2\ri\sqrt{n}\,P \mp 2\sqrt{n+2}\,Q}}\ ,
\eea
\bea\label{amplita}
&&A_{XX}+A_{YY} =
 - {{2\ri\sqrt{n}\,P}\over{1+n-2\ri\sqrt{n}\,P}}\times \\ 
\nonumber  && \qquad {{1 +
4n^2\,P^2 + 4(n+2)^2\,Q^2}\over{(1-2\ri\sqrt{n}\,P -2\sqrt{n+2}\,Q)(1 -
2\ri\sqrt{n}\,P + 2\sqrt{n+2}\,Q)}}\ .
\eea
The hairpin boundary state is a superposition of these Ishibashi
states,
\bea\label{boundst}
|\,  B\,  \rangle_{\subset} = \int\ \rd P\rd Q\ 
B (P,Q)\ |\,
I_{\bf P}\, \rangle\,,
\eea
where $B(P,Q)$ coincides with the vacuum overlap of the hairpin boundary
state, $B(P,Q)=\langle\, {\bf  P}\, |\,
B\, \rangle_{\subset}$. Its exact form can be figured out from the 
anticipated singularity structure,
\bea\label{bounhair}
&&B (P,Q) = g_D^2\ \Big({\kappa\over n}\Big)^{\ri  {P\over \sqrt n}
}\times\\ \nonumber &&
\ \ \ \ {{\sqrt{n}\ \Gamma(-\ri\sqrt{n}\,P)\, \Gamma(1 -
{{\ri P}/\sqrt{n}})}\over{\Gamma\big({1\over 2} -
 \sqrt{n+2}\ {Q\over 2}- \ri\sqrt{n}\ {P\over 2})\,
\Gamma\big({1\over 2} +  \sqrt{n+2}\ 
{Q\over 2}-  \ri\sqrt{n}\ {P\over 2}
\big)}}\ ,
\eea
and then the overall expressions\ \eqref{boundst},$\,$\eqref{bounhair}\ can 
be checked against the continuous version of  Cardy's 
consistency condition\ \cite{Cardy}.
Here we 
skip this part in view 
of the recent paper\ \cite{Schomerus}\ which 
seems to contain closely related 
calculation for the case of $D1$-brane in
the ``cigar'' sigma-model.

Note that the $W$-symmetry \eqref{wcftinv} of the hairpin model, 
together with its defining relations\ \eqref{comscreen},
strongly suggests the possibility of ``dual''
description of the hairpin model, 
similar to the duality between the
conformal cigar sigma-model 
and the ``sine-Liouville'' model\ \cite{zapiski, KK}\,.
Indeed, such dual description is possible, but we did not
yet elaborate all its details; 
we hope to come back to this issue
elsewhere. 
                                                                               
As was mentioned above, 
the right hairpin is the $X\to -X$ reflection of the
left hairpin described above. 
In particular, for the boundary 
state\ $|\,B\, \rangle_{\supset}$\ of the right
hairpin, the overlap amplitude $\langle\, {\bf P}\, |\,
B\, \rangle_{\supset}$ equals $B(-P,Q)$, and the whole
boundary state of the right hairpin 
satisfies the $W$-symmetry conditions
\bea\label{skxsu}
\big[\, z^s\, W_{s}^{(\supset)} (z) 
- {\bar z}^s\,{\bar W}_{s}^{(\supset)}
  ({\bar z})\, \big]_{|z|=R} \ |\, B\,  \rangle_{\supset} = 0
\, ,
\eea
where $W_{s}^{(\supset)}$ are 
the $X$-reflected versions of the $W$-currents
described above (i.e. 
$W_{s}^{(\supset)}$ are obtained from $W_s$ by repacing 
$X\to -X$ in their 
expressions in terms of the fields $(X,Y)$). Since in 
general the $W$-currents 
do not have symmetry with respect to these reflections, 
$\big\{W_s\big\}$ and
$\big\{W_{s}^{(\supset)}\big\}$, as the sets of operators in 
${\cal F}_{\bf P}$, are
different. These sets 
have some intersection though, as we will explain in
the next subsection, and further exploit in  Section\ \ref{secfive2}.

\subsection{More on the $W$-algebra}\label{secfour2}

Here we want to make few statements on the structure of the
$W$-algebra of the hairpin model.
                                                                               
The easiest way to generate 
the higher-spin $W$-currents is to observe that
the ``screening operators'' associated with the exponentials 
\eqref{screen}\ commute with
the parafermionic currents
\bea\label{parasha}
&\Psi (z)& = \big(\, \sqrt{n+2}\ \partial_z Y  + \ri
\sqrt{n}\ \partial_z X \, \big)\ 
\re^{{{2\ri}\over\sqrt{n+2}}\,Y_R(z)}\ , \\
\nonumber
&\Psi^{*}(z)& =
\big(\, \sqrt{n+2}\ \partial_z Y  - 
\ri\sqrt{n}\ \partial_z X \, \big)
\re^{-{{2\ri}\over\sqrt{n+2}}\,Y_R(z)}\ ,
\eea
in the same sense as the $W$-currents do, i.e.
\bea\label{psivertex}
\oint_{z}\rd w\ \Psi(z)\,{\cal V}_{\pm}(w)  = 0\,,\ \  \qquad
\oint_{z}\rd w\ \Psi^{*}(z)\,{\cal V}_{\pm}(w)  = 0\,.
\eea
In \eqref{parasha}
the $Y_R(z)$ in the exponential stands for the holomorphic
part of the local 
field $Y(z,{\bar z}) = Y_R(z) +  Y_L({\bar z})$, and
therefore the 
fields \eqref{parasha} are not local -- they extend the 
notion of the ${\Bbb Z}_k$ 
parafermions of \cite{ZamFat}\ to 
non-integer $k=-n-2$. Nonetheless, both $\Psi$ and
$\Psi^*$ are local with respect to the exponentials\ \eqref{screen},
hence the
integration contour in \eqref{psivertex} -- a 
small contour around $z$ -- is indeed
a closed one. It follows from \eqref{psivertex} that 
all the fields generated by the
operator product expansion of $\Psi(z)\Psi^{*}(w)$ 
satisfy Eqs.\eqref{comscreen}\ \cite{Bakas}. Thus
we have
\bea\label{psiope}
&&\Psi(z)\Psi^{*}(0) =  z^{-{2\over n+2}} \ 
\Big\{\, -{n+2\over z^{2}}
-{n\over 2}\  \big(T(z)+T(0)\big)
+\\ && \nonumber
{ \ri\, z
\over 
\sqrt{n+2}}\ \, \big(\, W_3(z)+W_3(0)\, \big)-
  { \ri\, z^2\over
4\sqrt{n+2}}\  \big(\, W_4(z)+W_4(0)\, \big)+\ldots\, \Big\}\, ,
\eea
where $T$ and $W_3$ are the same as in\ \eqref{energymom}\ and 
\eqref{wwwcurrent}, and the higher-order
terms involve the higher-spin $W$-currents.

Explicit calculation shows that the current $W_4$ appearing in \
\eqref{psiope} can be written as
\bea\label{w4}
W_4 = W_{4}^{({\rm sym})}+\partial_z{\cal O}_3 +a\ T^2+b\ \partial^2_z T\, ,
\eea
where
\bea\label{w4sym}
\nonumber
W_{4}^{({\rm sym})} &=&(4n^2+9n+4)\ (\partial^2_z X)^2 +
(4n^2+7n+2)\ (\partial^2_z Y)^2+\\ 
\nonumber &&
 n\, (3n+4)\ (\partial_z X)^4
+(n+2)(3n+2)\ (\partial_z Y)^4+\\ &&
6\, n( n+2)\ (\partial_z X)^2(\partial_z Y)^2\ ,
\eea
and ${\cal O}_3$ is some local field. 
The last two terms in\ \eqref{w4}\ are irrelevant -- they
represent the 
ambiguity in adding composites of the lower $W$-currents -- 
and exact values of the constants $a,b$ are not important here. 
Explicit form of 
the field ${\cal O}_3$ is not important for the present discussion
either. Important is the 
symmetry of $W_{4}^{({\rm sym})}$ with respect to the $X\to
-X$ reflection. 

Similar property can be observed for the next few
$W$-currents of even spins. Using a freedom in adding
derivatives and 
composite fields built from the lower-spin $W$-currents, the
fields $W_{2k}$ can be brought to the form
\bea\label{hahax}
W_{2k} = W_{2k}^{({\rm sym})} + \partial_z {\cal O}_{2k-1}\, ,
\eea
where the expressions for $W_{2k}^{({\rm sym})}$ are even with 
respect to the
reflection $X\to -X$. We believe this statement is valid 
for all even-spin $W$-currents. 
Since the ``right
hairpin'' $W$-currents $W_{s}^{(\supset)}$ 
in\ \eqref{skxsu}\ are $X\to -X$
reflections of the $W_s$, 
this property can be restated as follows: the
even-spin $W$-currents can be defined in such a way that
\bea\label{jsysysu}
W_{2k}^{(\supset)} = W_{2k}^{(\subset )}
\qquad {\rm mod} \qquad \partial_z(\ldots)\ ,
\eea
where the notation $W^{(\subset )}_{s}\equiv W_{s}$ is 
used to stress that the
currents $W_s$ described in this 
section generate the symmetries of the 
left hairpin. 

\subsection{Short-distance expansion of the paperclip}\label{secfour3}

Qualitative discussion at 
the beginning of this section suggests that the 
hairpin overlap  amplitude $B (P,Q)$
controls the $\kappa\to 0$ asymptotic of the paperclip partition
function $Z(\, P,Q\, |\, \kappa\, )$. More precisely, $B(P,Q)$ gives
right asymptotic form of $Z(\,P,Q\,|\,\kappa\,)$ 
when $P$ lays in the upper
half-plane $\Im m \,P > 0$, and one has to take $B(-P,Q)$ to describe
the asymptotic of $Z(\, P,Q\, |\, \kappa\, )$ when $P$ is 
in the lower
half-plane. Since $B(P,Q)$ vanishes in the limit $R\to 0$ if $P$ is
taken in the ``wrong'' half-plane (note the factor $\textstyle{
\kappa^{\ri {P\over \sqrt{n}}}}$
in \eqref{bounhair}), this in turn suggests 
that the overall $R\to 0$ asymptotic
of the partition function is correctly expressed by the sum
\bea\label{jauy}
Z(\, P,Q\, |\, \kappa\, )
\big|_{\kappa\to 0}\to B (P,Q) + B ( -P,Q)\, .
\eea

What can be said about corrections 
to this leading asymptotic? Let us
assume that $\kappa$ is 
small but not asymptotically small; that means that
the $X$-size of the paperclip in Fig.1 is large but finite. Also
assume that $\Im m \,P$ is positive. Then the functional integral
\eqref{path}\ is still dominated by 
the field configurations $(X,Y)$ with $X$
close to the left  end of 
the paperclip, with some fluctuations
towards its right end. One then expects to have two kinds of
corrections. When the fluctuation is sufficiently small, the
functional integral ``feels'' only the small deviations of the shape
of the paperclip \eqref{shape} from the shape \eqref{hairshpe} of 
the left hairpin.
This leads to the perturbative corrections to the
above leading $\kappa\to 0$ asymptotic; 
simple dimensional analysis shows
that these corrections behave as integer powers of 
$\textstyle{\kappa^{2\over n}}$, or
equivalently, as integer powers of ${ r}^2$. 
This part of the
structure is already explicit in the semiclassical
expression\ \eqref{newsemi}. 
Also, there are large fluctuations, sensitive to the
fact that the right end of the paperclip allows for the passage from
one leg of the left hairpin to another. Important example of those
are the instanton fluctuations discussed in 
Section\ \ref{secthree2}. The instantons
generate powers of ${  r}^n$ or, 
equivalently, integer powers
of $\kappa$.

This structure is 
neatly captured by the form
\bea\label{structure}
Z(\, P,Q\, |\, \kappa\, ) = 
B(P,Q)\  F(\, P,Q\, |\, \kappa\, ) + 
B(-P,Q)\  F(- P,Q\, |\, \kappa\, )\,,
\eea
where the function $ F
(\, P,Q\, |\, \kappa\, )$, apart from the overall factor $\kappa^{2\kappa}$
(see Section\ \ref{sectwo2}), is a double power series in
integer powers 
of $\kappa^{2\over n}$ and $\kappa$,
\bea\label{doubleser}
F(\,P,Q\,|\,\kappa\,) =\kappa^{2\kappa}\ 
\sum_{i,j=0}^{\infty}\,\,f_{i,j}(\,P,Q\,)\,\kappa^{i +{{2j}\over n}}\ ,
\eea
with $f_{0,0} =1$.  The exact splitting into two
terms in\ \eqref{structure}\ is 
not easy to justify on general grounds, but can be
supported by the following arguments. First, recall that such
splitting in the semiclassical 
expression\ \eqref{newsemi}\ corresponds to
isolating contributions 
from two saddle points, the ``right'' one near
the left end of the paperclip, 
and the ``wrong'' one near its right
end (We are still assuming 
that $\Im m\,P > 0$; otherwise the above
qualifications reverse. They have nothing to do with the authors
political standings.). Similar splitting occurs in the semiclassical
one-instanton contribution (see Section\ \ref{secthree2}), 
for the same reason. More
importantly, the full 
expression has to take care of the poles of the
factor $B(P,Q)$ -- recall 
that $Z(\, P,Q\, |\, \kappa\, )$ must be an entire function of
$P$. The expression\ \eqref{structure}\  is 
the simplest form fit for this job.
The poles of $B(P,Q)$ are located at the points in the lower half
of the $P$-plane 
where $\ri\, 
\sqrt{n}\,P$ or $\ri\, \textstyle{P\over \sqrt{n}}$ take non-negative
integer values. At this points the factor $\textstyle{
\kappa^{\ri {P\over \sqrt{n}}}}$ in $B(P,Q)$
``resonates'' with certain terms of the expansion \eqref{doubleser} 
in the second term in\ \eqref{structure}. Therefore, the
poles of the factor $B(P,Q)$ in the first term can (and must) be
cancelled by poles in appropriate terms of the expansion of the
second term. For the ``perturbative'' poles 
at $\ri\, \sqrt{n}\,P = 0, 1, 2\ldots $
this mechanism is evident in the semiclassical expression 
\eqref{newsemi}, and it can be
confirmed for the first of the ``instanton'' pole at $\ri\, {\textstyle{
P\over \sqrt{n}}} =
1$ by the form of the one-instanton correction \eqref{zinst}\,.
The form similar to \ \eqref{structure}, 
together with this mechanism of the pole
cancellation,  was previously observed in the boundary sinh-Gordon
model \ \cite{AlZ}.

\section{ Integrable Paperclip}\label{secfive}

In this section we would like to argue that the paperclip model is
integrable. We will put the arguments in turn below, but let us first
formulate the problem for our case, where the bulk theory is free CFT and
all the interactions occur at the boundary.

First, we need to have integrable bulk theory. This is not a problem
since the bulk theory is free, as described by the action\ \eqref{action}.
The free
theory certainly has an infinite set of commuting local integrals of
motion. Actually, it has several such sets, as we are going to see
below. The existence is clear, but to the best of our knowledge such
sets have never been completely classified. The problem is relatively easy
to address, and we believe it still can have surprises in store.

For the sake of this discussion it is convenient to change to the
cylindrical coordinates $(v,{\bar v}) = (\tau + \ri\sigma,\,
\tau-\ri\sigma)$ related to the disk coordinates
$(z,{\bar z})$ through the logarithmic map
\bea\label{sxut}
z/R = \re^{\ri {v/ R}}\,, \qquad\  {\bar z}/ R =
\re^{ -\ri {\bar
v}/ R}\,,
\eea
because relevant integrals of motion are going to be homogeneous in this
frame. The disk $|z|<R$ becomes a semi-infinite cylinder $\tau\equiv
\tau+2\pi R$, $\sigma > 0$. The action still has the form\ \eqref{action},
with the
derivatives $\partial_{z}, \partial_{\bar z}$ replaced by
$\partial_{v}, \partial_{\bar v}$, and the integration limits changed
accordingly. The fields ${\bf X}$ still obey the constraint
\eqref{shape}\ at the boundary, which is now placed at $\sigma = 0$.
The effect
of the exponential insertion in\ \eqref{exponent} is
accounted for by imposing an
asymptotic condition on the behavior of the field ${\bf X}$ at the
infinite end of the cylinder: ${\bf X} \to  \textstyle{\ri\, 
{\sigma\over R}}
\
{\bf P}$ as
$\sigma \to +\infty$ ($\ri\, P$ real). As usual, the bulk equations of
motion
$\partial{\bar\partial}{\bf X} = 0$ suggest
that $\partial {\bf X}$ is a holomorphic field, and hence
any composite field built from $\partial {\bf X}$ and the higher
derivatives is a holomorphic field as well\footnote{Here and below in
this
section $\partial$ and ${\bar\partial}$ will stand for the derivatives
over the cylindrical coordinates, $\partial =\partial_v$
and ${\bar\partial} = \partial_{\bar v}$.}. We will generally denote
such polynomial as $P(v)$. Assuming that the coordinate $\sigma$ along
the cylinder is interpreted as the Euclidean time, the integral
${\Bbb I}[P] = \oint \textstyle{{\rm  d} v\over
2\pi}\, P(v)$ over closed contour around the cylinder
gives rise to a conserved charge of the bulk theory.

Now, let's assume that $P_{s+1} (v)$ is an infinite sequence of the
polynomials,
such that all the associated charges commute, i.e.
$\big[\,{\Bbb I}_s\,,\,{\Bbb I}_{s'}\,\big] = 0$, where ${\Bbb I}_s \equiv
{\Bbb I}[P_{s+1}]$ (we will
come back to the question of what it takes to have such a sequence).
Also, let ${\bar P}_{s+1} ({\bar v})$ be the corresponding sequence of
``left-moving'' currents, with ${\bar\partial}$ replacing
${\partial}$, and ${\bar {\Bbb I}}_s \equiv {\bar {\Bbb I}}[{\bar P}_{s+1}]
= \oint \textstyle{{\rm  d} {\bar v}\over 2\pi} \,{\bar P}_{s+1}({\bar v})$
are the ``left-moving'' charges; the ${\bar {\Bbb I}}_{s'}$
also commute among
themselves and with all ${\Bbb I}_s$. The theory with the boundary at
$\sigma =0$
is integrable if the boundary state $|\, B\, \rangle$ satisfies the
equations \eqref{integrable} for all members of the sequence.
Roughly speaking, the meaning is that the boundary
neither emits nor absorbs any amount of the combined charges $
{\Bbb I}_s - {\bar {\Bbb I}}_s$. The equations\ \eqref{integrable}\ are
the property of very
special ``integrable'' boundary conditions, which have to be such that
the differences $P_{s+1} (v) - {\bar P}_{s+1} ({\bar v})$, when specified
to
the boundary $\sigma=0$, reduce to total derivatives
$\partial_{\tau}(\ldots)$\,\cite{gz}\,.

With this understanding, there are in principle two
opposite routes of approach to the integrable boundary interactions.
One is in direct analysis of a given boundary interaction, having in
mind proving (or disproving) that the system of local integrals
satisfying\ \eqref{integrable}\ exists. Another is to start with the
search for
suitable system of commuting integrals of motion
$\{ {\Bbb I}_s\} $; once that is
found, one can try to identify local boundary condition which is
compatible
with  Eqs.\eqref{integrable}. In the next section we describe what is
likely to be the commuting sequence $\{{\Bbb I}_s\}$ associated with the
paperclip model.

\subsection{ Paperclip integrals} \label{secfive1}

The polynomials $P(v)$ can be classified according to
their spins, which coincide with their scale dimensions and equal
to the total number of derivatives $\partial$ involved. Below we assume that
the subscript $s+1$ indicates the spin of the
field $P_{s+1}(v)$. We will look for a sequence of polynomials of even
spins,
$\{P_2 (v),\, P_4 (v), \dots \}$, such that all the
operators
${\Bbb I}_s = \oint \textstyle{ {\rm  d}  v
\over 2\pi}\,  \,P_{s+1}(v)$ commute among themselves. Having in
mind the paperclip model, we assume that $P_2=T$, where
$T$  coincides with the
energy-momentum tensor of the bulk CFT, i.e.
\bea\label{Ione}
{\Bbb I}_1 = -\oint {\rd v\over 2\pi}\,
\big[\, \big(\partial X\big)^2 + \big(\partial
Y\big)^2\, \big]\,,
\eea
so that the difference ${\Bbb I}_1 - {\bar {\Bbb I}}_1$
coincides with the momentum in the $\tau$-direction. We also assume  that
all
the higher $P_{s}$ respect obvious symmetries of the paperclip
\eqref{shape}, i.e. they are symmetric with respect to both the
reflections
$X\to -X$ and $Y \to - Y$. Then one can try generic polynomials with these
symmetries for $P_4(v),\,  P_6(v), \ldots$
with the coefficients to be determined
from the commutativity
conditions $[\, {\Bbb I}_s\, ,\, {\Bbb I}_{s'}\, ] =0$.
The equations $[\, {\Bbb I}_1\,  ,\, {\Bbb I}_s\, ]=0$
are satisfied identically for any polynomials $P_{s+1}(v)$ with constant
(i.e. $v$-independent) coefficients. But the first nontrivial equation
$[\, {\Bbb I}_3\,  ,\, {\Bbb  I}_5\, ]=0$
turns out to be surprisingly rigid. Besides having few
``trivial'' solutions\footnote{The ``trivial'' solutions are as follows:
One
is $P_4 = T^2, \, P_6 = T^3 - \textstyle{1\over 3}\,
\big(\partial T\big)^2$ (all composites here are in terms of the
Virasoro OPE);
it corresponds to the ``KdV series'' in which all the currents $P_{s}$
are
built from the $c=2$ Virasoro generator $T$ 
(see Ref.\cite{blz} for details). The other
solutions correspond to the KdV series with $c=1$ involving only one
component
of the field ${\bf X}$, either $X$, or $Y$, or one of the combinations $X\pm
Y$. We call these solutions ``trivial'' because their existence was
assured upfront.}, it fixes uniquely a one-parameter set
of solutions. Using suggestive notation $n$ for the parameter, the
corresponding integral ${\Bbb I}_3$ is written as
\bea\label{sisf} \nonumber
{\Bbb I}_3 &=& \oint{\rd v\over 2\pi}
\,\Big[\, {n\over{6(3n+2)}}\,\big(\partial X\big)^4 +
{n(n+2)\over{(3n+2)(3n+4)}}\, 
\big(\partial X\big)^2 \big(\partial Y\big)^2 +\\ 
&&
{{n+2}\over{6(3n+4)}}\,\big(\partial Y\big)^4 
+ {{4n^2 + 9n + 4}\over{6(3n+2)(3n+4)}}\,
\big(\partial^2 X\big)^2 +\\ \nonumber 
&&{{4n^2 + 7n + 2}\over{6(3n+2)(3n+4)}}
\,\big(\partial^2 Y\big)^2\, \Big]\, ,
\eea
(the expression for ${\Bbb I}_5$
is too long to be comfortably written down here
and we present it in Appendix). The integral \eqref{sisf}\, is a
special
case of the first nontrivial integral of motion written down in
\cite{fatey1}
in relation with the sausage model\ \cite{sausage}\ in  dual representation. 
Once ${\Bbb
I}_3$ is fixed as in
\eqref{sisf},
the equations $\Bbb[\,
{\Bbb I}_3\, ,\, {\Bbb  I}_{s}\, ]=0$
with $s=7,\, 9,\,11\ldots $ can be studied spin by
spin. We have done that for $s=7$ and $s=9$; in these cases the above
commutativity condition fixes the operators ${\Bbb I}_s$ uniquely up to
the
normalizations. We take this as a strong indication
that an infinite sequence of commuting integrals $\{
{\Bbb I}_s\,;\ s=1,3,\ldots, 2k-1\ldots\} $ exists, whose first
representatives
are the operators
\eqref{Ione},$\ $\eqref{sisf}. 
Our statement here is that the expression
\eqref{sisf} is essentially unique, once one demands it to be a member of
the commuting sequence $\{{\Bbb I}_{2k-1}\}$.

The commuting sequence $\{{\Bbb I}_{2k-1}\}$ whose first representatives are
\eqref{Ione} and \eqref{sisf} is the unique candidate for the system of
local
integrals of the paperclip model; anticipating direct relationship, we
will refer to it as the ``paperclip
sequence''. Note that besides the obvious reflection symmetries, 
Eq.\eqref{sisf} \ (as well as the expressions for the higher integrals
${\Bbb I}_5, \, {\Bbb  I}_7,\ldots $) respects  rather
subtle formal symmetry of the paperclip equation\ \eqref{shape},
\bea\label{jshjs}
X\leftrightarrow  Y\,, \qquad n \leftrightarrow - n-2\ .
\eea
Also, in the limit $n\to\infty$ the integral\ \eqref{sisf}\ (and
hence all the
integrals ${\Bbb I}_{2k-1}$ of this series)
becomes invariant with respect to orthogonal
rotations of the vector ${\bf X}$, in accord with the ``circular-brane''
limit
of the paperclip.

\subsection{Paperclip integrals from $W$-currents}\label{secfive2}

Important thing to note about the local integrals ${\Bbb I}_s$ is that the
expressions for their densities $P_{s+1}$ do not involve any particular
scale. On the other 
hand, as was discussed in  Section\ \ref{secfour}\ in the
short-distance limit $\kappa\to 0$ the paperclip becomes a composition of
the
left and right hairpins. Then, if one assumes that the paperclip boundary
state $|\,B\,\rangle$ satisfies the equations\ \eqref{integrable}, the
same
must be true for the boundary states of both left and right hairpins,
\bea\label{lhairint}
(\, {\Bbb I}_s - {\bar{\Bbb I}}_s\,)\ |\, B\,  \rangle_{\subset} = 0\,,
\eea
and
\bea\label{rhairint}
(\, {\Bbb I}_s - {\bar{\Bbb I}}_s\,)\ |\, B\,  \rangle_{\supset} = 0\,.
\eea
This is indeed the case, at least for the first few integrals:
the paperclip integrals ${\Bbb I}_s$ ``reside'' inside both the
$W$-algebras,
$W^{(\subset)}$ and $W^{(\supset)}$ of the left and the right hairpins,
and the equations \eqref{lhairint} and
\eqref{rhairint} follow from the equations\ \eqref{wcftinv}\ and
\eqref{skxsu}. The way it works\footnote{The arguments in this section
follow closely the ideas implicit in\ \cite{fatey1}.}
follows from the properties of the even-spin
$W$-currents mentioned in Section\ \ref{secfour2}. Using the cylindrical
coordinates
$(v,{\bar v})$, the $W$-symmetry equations can be written
as\footnote{The fact that
\eqref{wvlr}
are equivalent to
\eqref{wcftinv}\ and\ \eqref{skxsu}\ is obvious when 
one chooses $W_3,\, W_4,\, \ldots$
to be the
conformal primaries. It follows that these two forms are equivalent with
any
choice of  the $W$-currents.}
\bea\label{wvlr}
&&\big[\,W_{s}^{(\subset)}(v) -
{\bar W}_{s}^{(\subset)}({\bar v})\,\big]_{v ={\bar v}}
\ | \,B\,\rangle_{\subset} = 0\,,\\ \nonumber
&&\big[\,W_{s}^{(\supset)}(v) -
{\bar W}_{s}^{(\supset)}({\bar v})\,\big]_{v ={\bar v}}
\ | \,B\,\rangle_{\supset} = 0\,.
\eea
Let us take  even $s=2k$ and assume 
that the currents $W_{2k}$ are chosen
in
such a way that  Eq.\eqref{jsysysu}\ holds. Then, integrating
\eqref{wvlr}
around the cylinder one arrives at 
Eqs.\eqref{lhairint},$\,$\eqref{rhairint} \ with
operators $ {\Bbb I}_{2k-1} 
=\oint{{ {\rm  d} v}\over{2\pi}}\,W_{2k}^{({\rm sym})}(v)$. This argument
implies that the densities $P_{2k}(v)$ of the paperclip integrals
coincide with $W_{2k}^{({\rm sym})}(v)$ up to normalizations.
For $k=2$ this fact is evident
when one compares\ \eqref{w4sym} with\ \eqref{sisf}; it can also be checked
directly for the next few integrals. If one recalls the definition of the
$W$-algebra through  Eq.\eqref{comscreen}, it also suggests that the
``paperclip'' sequence of local integrals $\{{\Bbb I}_{2k-1}\}$ can be
defined as the system of local operators which commute with four
``screening charges'', i.e.
\bea\label{screee}
&&\oint_{v} \,dw\,W_{2k}^{({\rm sym})}(v)\,{\cal V}_{\pm}^{(\subset)}(w) =
\partial_v\,(\ldots)\,,\\ \nonumber
&&\oint_{v} \,dw\,W_{2k}^{({\rm sym})}(v)\, {\cal V}_{\pm}^{(\supset)}(w) =
\partial_v\,(\ldots)\,,
\eea
where ${\cal V}_{\pm}^{(\subset)}=
{\cal V}_{\pm}$ are the exponentials \eqref{screen}
and
${\cal V}_{\pm}^{(\supset)}$ 
are their $X\to-X$ reflections\footnote{This in
turn
suggests a possibility of ``dual'' description of the paperclip model,
in
which the boundary constraint\ \eqref{shape}\ is replaced by the boundary
potential
\bea
\nonumber
{\cal A}_{B}^{(dual)} =
\oint_{|z|=R}\,{{ {\rm  d} z}\over{2\pi z}}\ \Big[\sum_{\varepsilon_1,
\varepsilon_2
  = \pm}\,A_{\varepsilon_1
\varepsilon_2}\,\re^{\varepsilon_1\sqrt{n}\,X_B +
  \varepsilon_2\,\ri\,\sqrt{n+2}\,{\tilde Y}_B}\, \Big]\,,
\eea
with ${\tilde Y}$ being the T-dual of $Y$.
Simple-minded idea of the coefficients
$A_{\varepsilon_1 \varepsilon_2}$ being just constants does not seem to
work
-- the coefficients must involve some additional boundary degrees of
freedom
akin to the quantum group generators in\ \cite{BLZ3}. 
We did not yet work out all
details of this dual theory, but we hope to come back to it in future.}.

\section{Infrared Paperclip}\label{secsix}

The $R\to \infty$ limit of the paperclip model
is governed by an infrared fixed point of the boundary RG flow,
whose nature is not exactly known upfront;
The infrared domain $R\to\infty$ is beyond access of the weak-coupling
methods, no matter how large $n$ is. When the RG time $t$ grows, the
paperclip in Fig.1 first shrinks in the $X$-direction, and then
collapses to something of the size $\sim 1$. At this point its curvature
is no longer small, and the semiclassical intuition based on the geometry
of
\eqref{shape} is not
too reliable. Nonetheless, the simplest idea, which corresponds
to a naive extrapolation of
\eqref{shape},$\,$\eqref{flow}, is that the paperclip ultimately shrinks
to a point, i.e. the
infrared fixed point is the Dirichlet boundary conditions
for both $X$ and $Y$:
\bea\label{endpoint}
{\bf X}_B = (0,0)\,.
\eea
Assuming this simplest scenario, and combining it with the integrability
of
the paperclip model, one can make predictions about the large-$\kappa$
expansion of $Z(\, P,Q\, |\, \kappa\, )$. Indeed, under this assumption the
boundary-state
operator ${\Bbb B}(\kappa)$,  Eq.\eqref{boperator}, is expected to have
the
asymptotic $\kappa\to\infty$ expansion \eqref{bexpan} in terms of the
paperclip integrals ${\Bbb I}_s$. Since the vacuum overlap amplitude
\eqref{exponent} coincides with the vacuum-vacuum matrix element of ${\Bbb
  B}(\kappa)$, this implies the $\kappa\to\infty$ expansion
\bea\label{irexpan}
\log Z(\, P,Q\, |\, \kappa\, ) \simeq\ \log(g_D^2)-
\sum_{k=1}^{\infty}\,C_{2k-1}\,\kappa^{1-2k}\ I_{2k-1}(P,Q) 
\eea
of the logarithm of the function \eqref{partit} in terms of the vacuum
eigenvalues of the paperclip integrals,
\bea\label{ivac}
{\Bbb I}_{2k-1}\ |\,{\bf P}\,\rangle =
R^{1-2k}\ I_{2k-1}(P,Q)\ |\,{\bf P}\,\rangle\,.
\eea
The term $\log(g_D^2)=- \log(\sqrt{2})$ in Eq.\eqref{irexpan}
is the boundary 
entropy of the Dirichlet boundary condition\ \eqref{endpoint}.
Despite the fact that the coefficients $C_{2k-1}$ are not known {\it a
  priori},  Eq.\eqref{irexpan} gives rather strong prediction because
all
the dependence on $(P,Q)$ in \eqref{irexpan} comes through the eigenvalues
$I_{2k-1} (P,Q)$. These eigenvalues are polynomials in $P^2$ and $Q^2$
which
can be computed explicitly when explicit form of the paperclip integrals
${\Bbb
  I}_{2k-1}$ is known. Thus, using \eqref{Ione}, \eqref{sisf} and
\eqref{hsdtpo}\ one finds
\bea\label{sjusyu}
&&I_{1}(P,Q) = {P^2\over 4} + {Q^2\over 4} -{1\over 12}\,,
\\ \nonumber
&&I_{3}(P,Q) = {{n\ P^4}\over{96(3n+2)}} +
{{n(n+2)\ P^2\,Q^2}\over{16(3n+2)(3n+4)}}
+ {{(n+2)\ Q^4}\over{96(3n+4)}} -
\\
\nonumber
&&{{n(2n+3)\ P^2}\over{48(3n+2)(3n+4)}}  -
{{(n+2)(2n+1)\ Q^2}\over{48(3n+2)(3n+4)}}  +
{{18n^2+36 n
+11}\over{ 1440(3n+2)(3n+4)}}\, ,
\eea
and
\bea\label{shyyl}
I_{5}(P,Q)&=&\frac{n^2\ P^6}{320(5n+2)(5n+4)}
  +\frac{(n+2)^2\ Q^6}{320(5n+6)(5n+8)}\nonumber\\
 &&+\frac{3n^2(n+2)\ P^4\, Q^2}{64(5n+2)(5n+4)(5n+6)}
  +\frac{3n(n+2)^2\ P^2\, Q^4}{64(5n+4)(5n+6)(5n+8)}\nonumber\\
 &&-\frac{n^2(3n+4)\ P^4}{64(5n+2)(5n+4)(5n+6)}
  -\frac{(n+2)^2(3n+2)\ Q^4}{64(5n+4)(5n+6)(5n+8)}\nonumber\\
&&-\frac{3n(n+2)(5n^2+10n+2)\ P^2\, Q^2}{32(5n+2)(5n+4)(5n+6)(5n+8)}
\nonumber\\
&&+\frac{n(76n^3+250n^2+225n+30)\ P^2}{320(5n+2)(5n+4)(5n+6)(5n+8)}\\
 &&+\frac{(n+2)(76n^3+206n^2+137n+28)\ Q^2}
    {320(5n+2)(5n+4)(5n+6)(5n+8)}\nonumber\\
&&-\frac{1420n^4+5680n^3+7385n^2+3410n+564}{20160(5n+2)(5n+4)(5n+6)(5n+8)}\ .
\nonumber
\eea

\section{Solvable Paperclip}\label{secseven}

In this section we propose an exact expression for the boundary
amplitude\ \eqref{exponent}. The expression is in terms of solutions of
special ordinary second-order differential equation. Similar expressions
are known in a number of integrable models of boundary interaction,
beginning
with the work of Dorey and Tateo \cite{toteo}. Our proposal extends
this
relation to the paperclip-brane model. In this case no proof is yet
available,
but we will show in this section that the proposed expression reproduces
all
the properties of the paperclip amplitude described above.

\subsection{Differential equation}\label{secseven1}

Consider the ordinary second order differential equation
\bea\label{diff}
\bigg[\, -{{\rd^2}\over{\rd x^2}}   -
p^2\, { {\re^x}\over{1+\re^x}} -
\big(q^2-{\textstyle{1\over 4}}\big)
\, {{\re^x}\over{(1+\re^x)^2}} + \kappa^2\,\big(1+
\re^x\big)^n\, \bigg]\ \Psi(x) = 0\,,
\eea
where $p$, $q$ are related to the components of ${\bf P} =  (P,Q)$ in
\eqref{exponent},
\bea\label{PQpq}
p={\textstyle{1\over 2}}\ \sqrt{n}\ \, P\, ,
\ \ \ \ q={\textstyle{1\over 2}}\  \sqrt{n+2}\  \, Q\, ,
\eea
and $\kappa$ is proportional to the radius of the disk, $\kappa =
E_{*}R$.
With this identification in mind, below we always assume that $\kappa$
is real and positive. In the semiclassical case $n\gg 1$ the parameters
$p$, $q$ here are the same as $p$, $q$ in\ \eqref{kl}, and $\kappa$ here
relates to the renormalized $r = r(R^{-1})$ in \eqref{shape} through 
Eq.\eqref{flow} within the two-loop accuracy\footnote{The higher-loop
terms
in \eqref{flow} are scheme-dependent, but it is plausible that a scheme
exists in which the curve \eqref{shape} is perturbatively exact,
provided
\eqref{flow} is modified as follows
$$
\kappa = (n+1)\ r^n\,(1-r^2)\,.
$$
With this modification, the symmetry transformation \eqref{jshjs} of the
equation \eqref{shape} leaves $\kappa$ invariant.}.

Let
$\Psi_{-}(x)$ be the solution of\ \eqref{diff}\ which decays when $x$ goes
to
$-\infty$ along the real axis, and $\Psi_{+}(x)$ be another solution
of\ \eqref{diff},
the one which decays at large positive $x$. We fix normalizations
of these two solutions as follows,
\bea
\label{psiassminus}
\Psi_{-} \to \re^{\kappa x} \qquad\qquad {\rm as} \qquad \quad x\to
-\infty\,,
\eea
and
\bea\label{psiassplus}
\Psi_{+} \to \exp\bigg\{-\big({\textstyle{n\over 4}}+\kappa\big)\,x -
\kappa\, \int_{0}^{\re^x }{\rd z\over z}\,
\big(\,  (1+z)^{n\over 2}-1\, \big)\, \bigg\}
\ \,  \,  {\rm as}\, \, \ x\to +\infty\, .
\eea
Let
\bea\label{wronskian}
W[\Psi_{+},\Psi_{-}]\equiv \Psi_{+}\, {\rd\over \rd x}\,  \Psi_{-} -
\Psi_{-}\, {\rd\over \rd x}\,  \Psi_{+}
\eea
be the Wronskian of these two solutions. Then, our proposal
for the function\ \eqref{sfactor}, normalized in accordance
with\ \eqref{fixamb}, is
\bea\label{exactz}
 Z(\, P,Q\, |\, \kappa\,) =
\sqrt{\pi\over 2\kappa}\ { (2\kappa)^{2\kappa}\, \re^{-2\kappa}
\over{\Gamma(1+2\kappa)}}\ \ W[\Psi_{+},\Psi_{-}]\, .
\eea
To make more clear the motivations behind this proposal let us discuss
some
properties of the solutions of the differential equation\ \eqref{diff}.

\subsection{Small $\kappa$ expansion}\label{secseven2}

The differential equation \eqref{diff} has
the form of one-dimensional Schr${\ddot{\rm o}}$dinger
equation with the potential $V(x)$ defined by the last three terms in the
brackets in \eqref{diff}. In our analysis below, it will be often
convenient to split the potential into two parts,
\bea\label{diffop}
V(x)=V_{-}(x) + V_{+}(x)\, ,
\eea
where
\bea\label{Uminus}
V_{-}(x) =  -{p^2}\ {{\re^x}\over{1+\re^x}} -
\big(q^2 -{\textstyle{1\over 4}}\big)\ {{\re^{x}}\over{(1+\re^x)^2}}\,,
\eea
and
\bea\label{Uplus}
V_{+}(x) = \kappa^2\ (1+\re^x )^n\ .
\eea
                                                                                          
When $\kappa$ goes to zero the potential in\ \eqref{diff}\ develops a wide
plateau at
\bea\label{plateau}
1 \ll x \ll {\textstyle{1\over n}}\ 
\log\big({\textstyle{1\over \kappa^{2}}}\big)\, ,
\eea
where its value is close to $-p^2$. In this domain each of the solutions
$\Psi_{+}(x)$ and $\Psi_{-}(x)$ is a combination of two plane waves,
\bea\label{waves}
\Psi_{\pm}(x) = D_{\pm}(p,q)\ \re^{+\ri p\,x} + D_{\pm}(-p,q)
\ \re^{-\ri p \,x}\,.
\eea
Obviously, the Wronskian\ \eqref{wronskian}\ is
written in terms of the coefficients
in\ \eqref{waves}\ as follows,
\bea\label{det}
W\big[\Psi_{+},\Psi_{-}\big]_{\kappa\to 0}
\to -2 \ri\,  p\,\big(D_{+}(p,q)\,D_{-}(-p,q)
- D_{+}(-p,q)\,D_{-}(p,q)\big).
\eea
Then\ \eqref{exactz}\ leads
to the $\kappa\to 0$ limit of the partition function of
the form\ \eqref{jauy}, with
\bea\label{bcc}
B(P,Q) = -2\ri\,  p\ \sqrt{{\pi}\over 2{\kappa}}\ D_{+}(p,q)\,D_{-}(-p,q)\,,
\eea
where $P,\, Q$ related to $p,\, q$ through Eqs.\eqref{PQpq}.
The amplitude $D_{-}(p,q)$ is easy to determine since for $x\ll
{\textstyle {1\over
  n}}\,\log\big({\textstyle{1\over \kappa^{2}}}\big)$ 
the term $V_{+}(x)$, Eq.\eqref{Uplus}, is
negligible. In fact, to handle properly the asymptotic
condition\ \eqref{psiassminus},
it is convenient to retain the constant part $\kappa^2$ of $V_{+}(x)$.
With
$V_{+}(x)$ replaced by $\kappa^2$,  Eq.\eqref{diff}\ reduces to the
Riemann differential equation, and one finds
\bea\label{hypergeom}
&&\Psi_{-}(x) =
\re^{\ri p\, x}\ (1+\re^{-x})^{\ri p-\kappa}\, \times\\
&&\qquad {}_2F_1\big(\, {\textstyle{1\over 2}}+\kappa+q
 - \ri\, p , {\textstyle{1\over 2}}+\kappa-q- \ri\, p,1+2\kappa\, ;\,
{\textstyle{1\over{1+\re^{-x}}}}\, \big)\nonumber
\eea
as the $\kappa\to 0$ approximation of exact $\Psi_{-}(x)$ in the domain
$x \ll {\textstyle {1\over n}}
\,\log\big({\textstyle{1\over \kappa^{2}}}\big)$. 
Note that the way\ \eqref{hypergeom}\ is
written, it
captures the correction $\sim\kappa$, but mistreats the terms $\sim
\kappa^2$
and higher. We will come back to this expansion later.
From\ \eqref{hypergeom},
\bea\label{cminus}
D_{-}(p,q) ={ \Gamma(2\ri\,p)
\over \Gamma\big({1\over 2} -
 q+ \ri\, p \big)
\Gamma\big({1\over 2} + q+\ri\, p \big)
}\, .
\eea
On the other hand,
when $x\gg 1$ the $V_{-}(x)$ part of the potential becomes a
constant, while $V_{+}(x)$ can be approximated by $\kappa^2\,\re^{nx}$. To
understand the quality of this approximation one can make a change of the
variable $x=x_0 + {\textstyle {2\over n}}\, y$,
where $x_0 = {2\over n}\,\log\big({\textstyle{n\over 2\kappa}}\big)$;
the equation\ \eqref{diff}\ then takes the form
\bea\label{difff}
\bigg[-{{\rd^2}\over{\rd y^2}} -
{{4p^2}\over n^2} + \re^{2y} + \delta V(y)\, \bigg]\,\Psi
=0\,,
\eea
where $\delta V \sim \kappa^{2\over n}$ as $\kappa\to 0$, and indeed
$\delta
V$ admits expansion in powers of $\kappa^{2\over n}$,
\bea\label{delu}
\delta V(y) = \Big({{2\kappa}\over n}\Big)^{2\over n}\ \Big(\, n\,\re^{2y}
+{ {{4p^2 - 4q^2 +1}\over{n^2}}}\, \Big)\ \re^{-{{2y}\over n}} +
O\big(\kappa^{4\over n}\big)\, .
\eea
Therefore for $x\gg 1$
\bea\label{sjuyy}
\Psi_{+}(x) = \sqrt{{4\kappa}\over{\pi n}}\ \, \re^{{\cal E}\kappa}\
\Big[\, K_{\ri {\textstyle
 {2p\over n} }} \big( {\textstyle {{2\kappa}\over
  n}\,\re^{{nx}\over 2}}\big) + O(\kappa^{2\over n})\, \Big]\, .
\eea
Here
$K_{\nu}(z)$ is the conventional Macdonald function, and
\bea\label{ssyypo}
{\cal E}=\gamma_E+\psi\big(-{\textstyle{n\over 2}}\big)\, ,
\eea
with $\gamma_E=0.5772\ldots$ being Euler's constant, and
$\psi(z)={\textstyle {{\rm d}\over {\rm d} z} \log\Gamma(z)}$.
The overall normalization in  Eq.\eqref{sjuyy}\ is
chosen to ensure the asymptotic form\ \eqref{psiassplus}.
Hence
\bea\label{sjdsudu}
D_{+}(p,q) = \sqrt{{\kappa}\over{\pi n}}\ \, \re^{{\cal E}\kappa}\
\Big({\kappa\over
  n}\Big)^{2\ri\,
{\textstyle {p\over n}
}}\ \Gamma\big(-2\ri\,  {\textstyle{{ p}\over{n}}}\big)\, .
\eea
Thus,  Eq.\eqref{bcc}\ reproduces the 
correct form\ \eqref{bounhair}\ of the hairpin boundary
amplitude $B(P,Q)$.
                                                                                          
When $\kappa$ is small but finite,
corrections to \eqref{waves}\ can be obtained
using perturbation theory.
In computing $\Psi_{-}(x)$ the term $V_{+}(x)$ is
treated as the perturbation, and corrections appear in the form of a power
series in $\kappa$. In fact, the leading correction $\sim\kappa$ is
already
contained in\ \eqref{hypergeom},
and higher powers of $\kappa$ are generated by
further iterations.
On the other hand, $\Psi_{+}(x)$ is computed from\ \eqref{difff}
with the term $\delta V(x)$ taken as the perturbation.
It is evident from
\eqref{delu}\ that corrections
to $\Psi_{+}(x)$ obtained through this perturbation
theory have the form of a power series in $\kappa^{2\over n}$.
As the result, 
the partition function computed according 
to\ \eqref{exactz}\ appears in the form\ \eqref{structure}
with $F(\, P,Q\, |\, \kappa\, )$\ being a double power 
series\ \eqref{doubleser}.
The low-order
terms can be easily obtained by
direct perturbative calculations outlined
above, for instance,
\bea\label{sxiusyx}
&&f_{1,0} =
{\rd\over{\rd\epsilon}}\, \log\bigg[\,
{2^{2\epsilon}
\re^{({\cal E}-2)\epsilon}\over
  \Gamma\big({\textstyle {1\over 2}}+\epsilon -q-\ri \, p\big)
\Gamma\big({\textstyle {1\over 2}}+\epsilon +q-\ri \, p\big) }
\, \bigg]_{\epsilon=0}=\\ && \nonumber
- \psi\big({\textstyle {1\over 2}} -q-\ri \, p\big)-
\psi\big({\textstyle {1\over 2}} +q-\ri \, p\big)+ \gamma_{E}+
\psi\big(-{\textstyle{n\over 2}}\big)+2\, \log(2 )-2\, ,
\eea
and
\bea\label{sxiusyxaa}
 f_{0,1}&=&-{\Gamma({1\over 2}+{1\over n})
\Gamma(1-{1\over n})\over \sqrt{\pi}}\ \Big({2\over n}\Big)^{{2\over n}-1}
\times \\ &&
\nonumber \Big(\, {n\over n-2}-
{4p^2-4q^2+1\over 2(4p^2+1)}\, \Big)\
{\Gamma(1-{1\over n}-2\ri \, {p\over n})\over
\Gamma({1\over n}-2\ri \, {p\over n})}\, .
\eea

\subsection{Semiclassical domain $\kappa \ll 1 \ll  n$}\label{secseven3}

When $\kappa$ is small but $n$ is large, so that $\kappa^{2\over n} \sim
1$,
all powers of $\kappa^{2\over n}$ in the above expansion have to be
collected. Let us assume that $r$ defined through \eqref{flow}\ is not too
close to 1, so that $n\,(1-r^2) \sim n$. This regime corresponds to the
semiclassical domain of the paperclip model 
considered in Section\ \ref{secthree}.
                                                                                          
First, let us consider the case when $p$ and
$q$ in \eqref{diff} are of the order of 1; it corresponds to the case of
``light'' insertion in \eqref{path}, with $(P,Q) \sim 
\textstyle{1\over \sqrt{n}}$.
In this regime the term
$V_{+}(x)$ in
the potential has the effect of a rigid wall at
some point $x_r$,
i.e. to the left from this point, for $x_r - x \gg \textstyle{1\over n}$, 
the potential
$V_{+}(x)$ is negligible, but to the right from  it grows very
fast, so that for $x>x_r$ the solution $\Psi_{+}(x)$ is essentially zero.
More precisely, when $x$ is below but close to $x_r$ the solution
$\Psi_{+}(x)$ is approximated by a linear function,
\bea\label{linear}
\Psi_{+}(x)  \approx \alpha_r\ (x_r - x)\,.
\eea
However, the position of the
wall $x_r$ and the slope $\alpha_r$ depend on $\kappa$
in a somewhat subtle way. Here it is useful to trade $\kappa$ for the
``running coupling constant'' $r$ defined through the equation
$\kappa = n\,r^n\,(1-r^2),$
which is identical to the perturbative RG flow equation of the paperclip,
Eq.\eqref{flow}. Then
\bea\label{jsysuy}
x_r = \log\big({\textstyle
{{1-r^2}\over {r^2}}}\big)+O({\textstyle{1\over n}}\big)\, ,
\eea
and
\bea\label{jssuy}
\alpha_r = \sqrt{{n\kappa}(1-r^2)\over\pi}\ \ 
\Big(1+O({\textstyle{1\over n}}\big)\, \Big)\, .
\eea
We will explain these relations shortly. Using them,
the Wronskian\ \eqref{wronskian}\ is
easily determined. Taking, for instance, any point $x$ close to
the left of
the wall where both\ \eqref{hypergeom}\ and \eqref{linear}\  apply,
one finds it
equal to $\alpha_r$ times the
expression \eqref{hypergeom}\ (with $\kappa$ set to zero)
evaluated at $x=x_r$ -- this
leads exactly to\ \eqref{semiclas}. Moreover, the
one-instanton contribution\ \eqref{zinst} is
also reproduced when one takes into account
the first-order term of the $\kappa$ expansion of  Eq.\eqref{hypergeom}.
                                                                                          
The case of ``heavy'' insertion can be handled similarly. The potential
$V_{+}(x)$ still can be treated as a wall at $x_r$, in the sense that at
$x_r - x \gg {\textstyle{1\over n}}$ its effect is negligible, but it 
dominates at $x - x_r \gg {\textstyle{1\over n}}$. 
Close to the wall's position, i.e. at
\bea\label{wall}
|x-x_r | \ll 1\,,
\eea
the part $V_{-}(x)$ is well approximated by the constant
\bea\label{sbsgtyt}
V_{-}(x) \approx V_{-}(x_r ) =
{\textstyle{1\over 4}}\ n\,(1-r^2 )^2\ P_{r}^2\ ,
\eea
where $P_r$ is exactly the expression \eqref{nub}. 
On the other hand, in the 
domain \eqref{wall} the part $V_{+}(x)$ behaves as the exponential
\bea\label{vpexp}
V_{+}(x)  \approx  n^2\,(1-r^2 )^2 \ \re^{2y}\,, \qquad y =
{\textstyle{1\over 2}}\  n\,(1-r^2) \ (x-x_r )\, .
\eea
Therefore, in the domain\ \eqref{wall}\ the solution $\Psi_{+}$ is
approximated by the Macdonald function, 
\bea\label{psiwall}
\Psi_{+}(x)  \approx {\textstyle {2\over{\sqrt{\pi}}}}\ \, r^{n\over 2}
\ K_{\ri \nu}\big(2\,\re^{y}\big)\,,
\eea
where $\nu = {\textstyle{P_r\over \sqrt{n}}}$\,. The
normalization factor in front of $K$ is fixed by 
 matching \eqref{psiwall} to
the asymptotic condition \eqref{psiassplus}\,. 
For $\nu \to 0$ and $-y \gg 1$ \eqref{psiwall} 
reduces to \eqref{linear}\,.  The Wronskian
\eqref{wronskian} can be evaluated in the domain 
${\textstyle{1\over n}} \ll x_r-x \ll 1$, where
both \eqref{hypergeom} and \eqref{psiwall} are valid; the result is
exactly \eqref{newsemi} with the extra factor \eqref{shsdy} added. 

\subsection{Large $\kappa$ expansion}\label{secseven4}

The case of large $\kappa$ is
expected to describe the infrared limit of the
paperclip, see Section\ \ref{secsix}.
For the differential equation\ \eqref{diff}\ it is the domain
of validity of the WKB approximation. Applying
the standard WKB iterational
scheme\ \cite{Landau}\ one finds for the Wronskian\ \eqref{wronskian},
\bea\label{wkbu}
\log W =\log(2\kappa)  +
\int_{-\infty}^{\infty}\rd x\,\bigg\{ \kappa \big(
{\cal P}(x)-{\cal P}_0 (x) \big) +
{1\over{8\kappa}}\,{{({\cal P}'(x))^2}
\over{{\cal P}^3(x)}} +\ldots  \bigg\} ,
\eea
where
${\cal P}(x)=\kappa^{-1}\, \sqrt{V(x)}$. The
subtraction term
\bea\label{shsylo}
{\cal P}_0 (x) = \big(1 + \re^x\big)^{n\over 2}
\eea
in the integrand derives from the asymptotic conditions\
\eqref{psiassplus}.
With the
potential $V(x)$ as in \eqref{diffop},
this generates asymptotic expansion of the
partition function\ \eqref{sfactor},
\bea\label{expint}
\log  Z(\, P,Q\, |\, \kappa) \simeq\ -{\textstyle{1\over 2}}\, \log(2)
-\sum_{k=1}^{\infty}\,\kappa^{1-2k}\ I_{2k-1}(P,Q)\,,
\eea
where $I_{2k-1}(P,Q)$ are polynomials of the variables $P^2$ and $Q^2$ of
the degree $2k$. Their highest-order terms follow from the first
term in the integrand in\  \eqref{wkbu},
\bea\label{ikhighest}
&&I_{2k-1}(P,Q) =
{{\Gamma(k-1/2)\,(k-1)!}\over{2^{2k+1}\sqrt{\pi}}}\,\times
\\ \nonumber
&&\ \ \ \sum_{l=0}^{k}\,{{n^{k-l}\,(n+2)^l}\over{l!\,(k-l)!}}\,
{{\Gamma\big((k-1/2)n +l\big)}\over{\Gamma\big((k-1/2)n+k+l\big)}}\,
\big(P^2\big)^{k-l}\,\big(Q^2\big)^l + \ldots \ ,
\eea
which is in agreement with the highest-order terms of the vacuum
eigenvalues \eqref{sjusyu},$\,$\eqref{shyyl}.
It is also straightforward to generate the
full polynomials evaluating the
integral\ \eqref{wkbu}\ order by order in $\kappa^{-2}$. This
calculation reproduces exactly the eigenvalues \eqref{sjusyu} and
\eqref{shyyl}, as
well as few higher-spin eigenvalues (we have verified exact agreement
up to $I_9$). Note that Eq.\eqref{ikhighest} fixes
normalizations of the operators ${\Bbb I}_{2k-1}$; with this
normalization all the coefficients $C_{2k-1}$ in \eqref{irexpan}
($C_s$ in \eqref{bexpan}) are equal to $1$.

\subsection{Circular brane limit}\label{secseven5}

As was mentioned in  
Introduction, in the limit when $n \to \infty$ with the
parameter $g^{-1} = n\,(1-r^2)$ kept fixed (so that $\kappa$ in 
\eqref{flow} has finite limit, 
$\kappa \to g^{-1}\, \re^{1\over{2g}}\, $), the
paperclip curve \eqref{shape} becomes a circle \eqref{circle}. It is
interesting to see how this limit is obtained in the differential equation
\eqref{diff}.

When $n$ goes to $\infty$ and $\kappa$ remains fixed, 
the term $V_{+}(x)$ in
the potential\ \eqref{diffop}\ becomes 
infinite at any finite $x$. Correct limit
is obtained by making first the shift of $x$,
\bea\label{logn}
x =y -\log n \, ,
\eea
so that $V_{+}(x) = \kappa^2\,\big(1+{1\over n}\,e^y\big)^n$. 
Then, taking the
limit $n\to\infty$ brings the equation \eqref{diff} to the form
\bea\label{circdiff}
\bigg[\, -{ {{{\rm d}^2}\over{{\rm d} y^2}} } 
- 
{{{\bf P}^2\over 4}} 
\ \re^y + \kappa^2\,\exp(\re^y)\, \bigg]
\ \Psi(y) = 0\,,
\eea
where ${\bf P}^2 = P^2 +Q^2$. Thus, our proposal\ \eqref{exactz}\ applies
directly to the circular brane model\ \eqref{circle}, with
$W[\Psi_{+},\Psi_{-}]$ being the Wronskian of two solutions of the
differential equation\ \eqref{circdiff}\ which are fixed by the asymptotic
conditions
\bea\label{minusas}
\Psi_{-}(y) \to \re^{\kappa\,y} \qquad {\rm as}\qquad y \to -\infty\,,
\eea
and
\bea\label{plusas}
\Psi_{+}(y) \to \exp\bigg\{-{\re^y\over 4}-\kappa y-\kappa
\int_0^{{1\over 2}\,  \re^y} {\rd z\over z}\ 
\big( \re^{z}-1 \big)\, \bigg\}\ \ \  {\rm as}\ \ \  y \to +\infty\, .
\eea
More details about the circular brane model, and more about the properties of
the differential equation\ \eqref{circdiff}, are 
presented in a separate paper\ \cite{SLAZ}.

\bigskip
\section*{Acknowledgments}

SLL and ABZ are grateful to Vladimir Fateev and  Alexei Zamolodchikov 
for valuable discussions. 
ABZ thanks Rainald Flume, Gunter von Gehlen,
and Vladimir Rittenberg for interest to this work, and for kind hospitality
extended to him at the Institute of Physics of the University of Bonn. 

\bigskip

\noindent
This research is supported
in part by DOE grant $\#$DE-FG02-96 ER 40949.
ABZ also acknowledges support
from Alexander von Humboldt Foundation.

\bigskip
\bigskip
\section{Appendix}
Here we present an explicit form for the density $P_6$
of the local  integral ${\Bbb I}_5=\oint{{\rm d} v\over 2\pi}\, P_6(v)$:
\bea\label{hsdtpo}
P_6&=& -\frac{n^2}{5 (5 n+2)(5 n+4)}\ \ (\pp X)^6\ \nonumber\\
 &&-\frac{(n+2)^2}{5 (5 n+6)(5 n+8)}\ \ (\pp Y)^6\nonumber\\
 &&-\frac{3 n^2(n+2)}{(5 n+2)(5 n+4)(5 n+6)}\ \ (\pp X)^4\,(\pp Y)^2
  \nonumber\\
 &&-\frac{3 n(n+2)^2}{(5 n+4)(5 n+6)(5 n+8)}\ \ (\pp X)^2\, (\pp Y)^4
  \nonumber\\
 &&-\frac{3 n (4n^2+7n+2)}{(5 n+2)(5 n+4)(5 n+6)}\
    \ (\pp X)^2\, (\pp^2X)^2\nonumber\\
 &&-\frac{3(n+2)(4n^2+9n+4)}{(5 n+4)(5 n+6)(5 n+8)}\
    \ (\pp Y)^2\, (\pp^2Y)^2\\
 &&-\frac{3n (n+2)(6n^2+13n+4)}{(5 n+2)(5 n+4)(5 n+6)(5 n+8)}\
    \ (\pp^2X)^2\, (\pp Y)^2\nonumber\\
 &&-\frac{3 n(n+2) (6n^2 +11n+2 )}{(5 n+2)(5 n+4)(5 n+6)(5 n+8)}\
    \ (\pp X)^2\, (\pp^2Y)^2\nonumber\\
 &&-\frac{12 n(n+2) (7n^2+14n+ 4)}{(5 n+2)(5 n+4)(5 n+6)(5 n+8)}\
      \ \pp X\pp^2 X\, \pp Y\, \pp^2Y\nonumber\\
 &&-\frac{131n^4+575n^3+860n^2+500n+96}{20 (5 n+2)(5 n+4)(5 n+6)(5 n+8)}
   \ \  (\pp^3X)^2\nonumber\\
 &&-\frac{131n^4+473n^3+554n^2+232n+32}{20 (5 n+2)(5 n+4)(5 n+6)(5 n+8)}
    \ \ (\pp^3Y)^2\ .\nonumber
\eea

\end{document}